\documentclass[a4paper,fleqn]{cas-dc}

\usepackage[numbers]{natbib}

\usepackage{grffile} 

\usepackage{tabularx} 

\def\tsc#1{\csdef{#1}{\textsc{\lowercase{#1}}\xspace}}
\tsc{WGM}
\tsc{QE}
\tsc{EP}
\tsc{PMS}
\tsc{BEC}
\tsc{DE}

\begin{document}
\let\WriteBookmarks\relax
\def\floatpagepagefraction{1}
\def\textpagefraction{.001}
\shorttitle{Optimizing implicit relaxation-zones}
\shortauthors{Perić et al.}

\title [mode = title]{Optimizing wave-generation and wave-damping in 3D-flow simulations with implicit relaxation-zones
}                      



\author[1]{Robinson Perić}
\ead{robinson.peric@tuhh.de}


\address[1]{Hamburg University of Technology (TUHH), Institute for Fluid Dynamics and Ship Theory (M8), Hamburg, Germany}

\author[2]{ Vuko Vuk$\mathbf{\check{\mathrm{c}}}$evi\'c}
\address[2]{SimScale GmbH, Riddlerstrasse 31b, Munich, Germany}

\author[1]{ Moustafa Abdel-Maksoud}

\author[3,4]{ Hrvoje Jasak}
\address[3]{University of Zagreb, Faculty of Mechanical Engineering and Naval Architecture, Ivana Lu$\check{\mathrm{c}}$i\'ca 5, Zagreb, Croatia}
\address[4]{Wikki Ltd, 459 Southbank House, SE1 7SJ, London, United Kingdom}









\begin{abstract}
In finite-volume-based flow-simulations with free-surface waves, wave reflections at the domain boundaries can cause substantial errors in the results and must therefore be minimized. This can be achieved via `implicit relaxation zones', but only if the relaxation zone's case-dependent parameters are optimized. This work proposes an analytical approach for optimizing these parameters. The analytical predictions are compared against results from 2D-flow simulations for different water depths, flow solvers, and relaxation functions, and against results from 3D-flow simulations with strongly wave-reflecting bodies subjected to nonlinear free-surface waves. The present results demonstrate that the proposed approach satisfactorily predicts both the optimum parameter settings and the upper-limit for the corresponding reflection coefficients $C_{\mathrm{R}}$. Simulation results for $C_{\mathrm{R}}$ were mostly below or equal to the analytical predictions, but never more than $3.4\%$ larger. Therefore, the proposed approach can be recommended for engineering practice. Furthermore, it is shown that implicit relaxation zones can be considered as a special-case of forcing zones, a family of approaches which includes among others absorbing layers, damping zones and sponge layers.  The commonalities and differences between these approaches are discussed, including to what extend the present findings are applicable to these other approaches and vice versa.
\end{abstract}

\begin{graphicalabstract}
 \includegraphics[width=\linewidth]{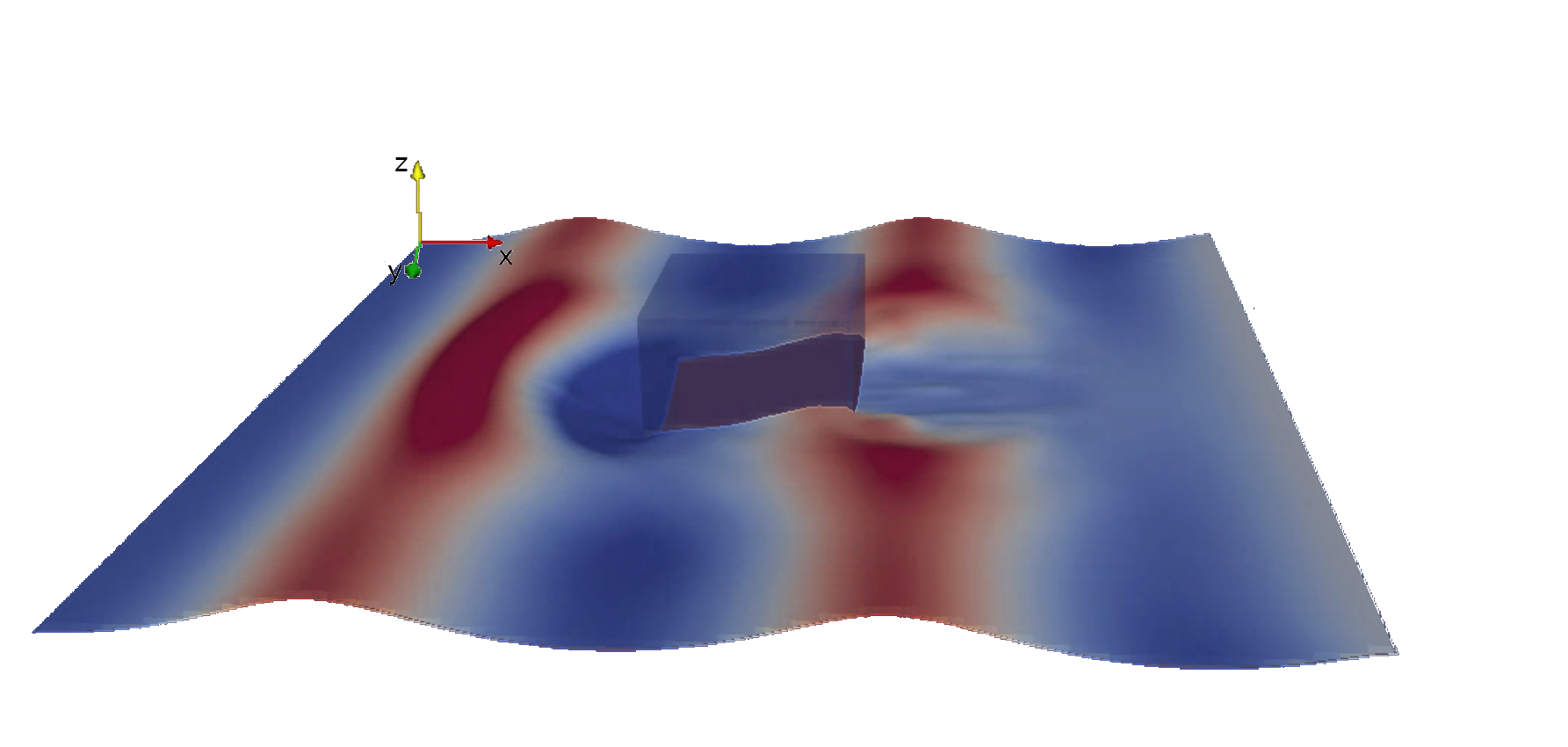} 
\end{graphicalabstract}

\begin{highlights}
\item An analytical approach is presented for optimizing wave-generation and wave-damping when using implicit relaxation zones
\item A computer program for optimizing the implicit relaxation zone based on the analytical approach was published as free software: \url{https://github.com/wave-absorbing-layers/relaxation-zones-for-free-surface-waves}
\item The approach predicts the optimum values of the relaxation zone's case-dependent parameters and closely estimates the upper-limit for the corresponding wave-reflection coefficients
\item Analytical predictions agree well with results from 2D- and 3D-flow simulations with strongly wave-reflecting bodies subjected to nonlinear free-surface waves
\end{highlights}


\maketitle

\section{Introduction}
\label{SECintro}
In finite-volume-based flow simulations with free-surface waves, accurate wave-generation and wave-damping at the domain boundaries is important.  Wave reflections at the boundaries of the computational domain  can cause substantial errors in the results and must therefore be minimized (cf. Mani, 2012; Perić and Abdel-Maksoud, 2016; Vyzikas et al., 2017; Windt et al., 2019). This can be achieved by implementing \textit{relaxation zones}, which gradually fade-out the simulated flow solution and blend-in a prescribed far-field wave solution near the domain boundaries. 

Relaxation zones can be subdivided into \textit{explicit} and \textit{implicit relaxation zones}, which are fundamentally different:

\textit{Explicit relaxation zones} (implemented e.g. in waves2Foam, cf. Jacobsen et al., 2012) modify the fields for volume fraction $\alpha$ and velocity $\mathbf{u}$ by replacing computed values $\phi_{\mathrm{computed}}$ by 
\begin{equation}
\phi = (1-b(\mathbf{x})) \phi_{\mathrm{target}} + b(\mathbf{x}) \phi_{\mathrm{computed}} \quad ,
\label{EQexplicitRelax}
\end{equation}
where $b(\mathbf{x})$ is a weighting function and $ \phi_{\mathrm{target}}$ is the target solution. This modification is performed in each time-step, e.g. prior to the solution of the pressure-velocity coupling  (cf. Jacobsen et al., 2012). 
Among the most influential implementations of explicit relaxation zones are Mayer et al. (1998), Madsen et al. (2003), Engsig-Karup et al. (2006), Fuhrman et al. (2006), Jacobsen et al. (2012); further references and comparison to other wave-generation and wave-damping approaches can be found e.g. in Schmitt and Elsaesser (2015), Windt et al., (2018, 2019) and Li et al., (2019).
In contrast to implicit relaxation zones, it is not directly apparent to which source terms in the governing equations the `explicit' manipulation of the flow field corresponds. Therefore, in this work the focus will be on implicit relaxation zones.

\textit{Implicit relaxation zones} (implemented e.g. in Naval Hydro Pack, cf. Jasak et al.; 2015, Vuk\v{c}evi\'{c} et al., 2016a, 2016b, 2017; Perić, 2019) introduce source terms in the governing equations  to blend, say a general transport equation $\mathcal{T}$ for transport quantity $\phi$, over to a reference solution $\phi_{\mathrm{ref}}$  via
\begin{equation}
\left( 1-b(\mathbf{x}) \right) \mathcal{T} + \frac{b(\mathbf{x})}{\tau} \mathcal{R} =  0 \quad ,
\label{EQrelaxprinciple}
\end{equation}
where $b(\mathbf{x})$ is a blending function such as  Eq. (\ref{EQblendpow}), $\mathcal{T}$ corresponds e.g. to the conservation equations for fluid momentum or volume fraction, and $\mathcal{R}$ corresponds to $\int_{V}  \left( \phi - \phi_{\mathrm{ref}} \right) \ \mathrm{d}V$.

The main problem with both explicit and implicit relaxation zones is that they  provide satisfactory wave-generation and wave-damping \textit{only if their case-dependent parameters are optimized}. However, how to optimize these case-dependent parameters before performing the flow-simulations has not been shown so far (cf. Miquel et al., 2018; Chen et al., 2019; Choi et al., 2020; Higuera, 2020). Thus at present, is it common practice to use either default or trial-and-error-based values for these parameters, which can lead to large errors in the results as will be demonstrated in this work.

\textit{Therefore, the first aim of this work is to present an analytical approach to optimize the case-dependent parameters of implicit relaxation zones}, so that these can be optimized before performing the flow simulation. 

Apart from relaxation zones, there exist various other approaches that generate and damp waves via domain-internal source terms. These approaches include `absorbing layers' (e.g. Wei et al., 1999), `damping zones' (e.g. Park et al., 1999, 2001), `dissipation zones' (Park et al., 1993), `numerical beaches' (e.g. Schmitt et al., 2019), `sponge layers' (e.g. Israeli and Orszag, 1981; Larsen and Dancy, 1983; Choi and Yoon, 2009) or the `Euler overlay method' (e.g. Kim et al., 2012). Recently, Perić (2019) showed that all these approaches can be formulated as special cases of a generic approach, called  \textit{forcing zones}. 
Forcing zones add source terms on the right-hand side of e.g. the conservation equations for velocity $u_{i}$ and volume fraction $\alpha$, to gradually force the flow towards a prescribed reference solution, $u_{i,\mathrm{ref}}$ and $\alpha_{\mathrm{ref}}$, near the domain boundaries:
\begin{align}
q_{i} = \int_{V} \rho \gamma  b(\mathbf{x}) (u_{i,\mathrm{ref}}-u_{i}) \ \mathrm{d}V \quad ,
\label{EQnavier_stokes}
\end{align}
\begin{equation}
q_{\mathrm{\alpha}} = \int_V \gamma   b(\mathbf{x}) \left( \alpha_{\mathrm{ref}} -\alpha \right) \ \mathrm{d}V \quad ,
\label{EQtransport_alpha}
\end{equation}
with volume $V$ and fluid density $\rho$. The case-dependent parameters of forcing zones are the zone thickness $x_{\mathrm{d}}$, the forcing strength $\gamma$, which regulates the source-term magnitude, and the blending function $b(\mathbf{x})$, which regulates how the source-term magnitude varies within the zone. The optimum values of these parameters can be determined analytically as shown by Perić and Abdel-Maksoud (2018). 
However, so far it is not known whether or to which extend the findings on the optimization of forcing zones are applicable to  relaxation zones.

\textit{Therefore, the second aim of this work is to show how implicit relaxation zones are related to forcing zones}, and to what extend findings obtained for forcing zones are applicable to implicit relaxation zones and vice versa.

Section \ref{SECgoveq} describes the governing equation for free-surface flows with implicit relaxation zones. Section \ref{SECcomputeCR} describes how to compute reflection coefficients, which are used to quantify how well waves are absorbed in the relaxation zone.
Section \ref{SECtheory} presents the analytical approach for optimizing the implicit relaxation zone. Sections \ref{SEC2dflow} and \ref{SECfsrelaxres3d} compare the analytical predictions against results from 2D- and 3D-flow simulations based on the setup from Sect. \ref{SECfsrelaxsetup}. Section \ref{SECdiscuss} discusses the findings and the relation between forcing zones, explicit relaxation zones and implicit relaxation zones.

\section{Governing equations with implicit relaxation zones}
\label{SECgoveq}
The conservation equations for momentum and volume fraction take the form
\begin{align}
\left( 1-b(\mathbf{x}) \right) \bigg[ \frac{\mathrm{d}}{\mathrm{d} t} \int_{V} \rho u_{i} \ \mathrm{d}V 
+ \int_{S} \rho u_{i} (\textbf{u} - \textbf{u}_{g} ) \cdot \textbf{n} \ \mathrm{d}S  \nonumber \\ 
-\int_{S} (\tau_{ij}\textbf{i}_{j} - p\textbf{i}_{i}) \cdot \textbf{n} \ \mathrm{d}S 
- \int_{V} \rho \textbf{g}\cdot \textbf{i}_{i} \ \mathrm{d}V  \bigg] \nonumber \\
+  \frac{b(\mathbf{x})}{\tau} \bigg[ \int_{V} \rho \left( u_{i} - u_{i,\mathrm{ref}} \right) \ \mathrm{d}V  \bigg] =  0 
\quad ,
\label{EQnavier_stokesRelax}
\end{align}
\begin{align}
\left( 1-b(\mathbf{x}) \right) \bigg[ \frac{\rm d}{{\rm d} t} \int_{V} \alpha \ \mathrm{d}V + \int_{S} \alpha (\textbf{u} - \textbf{u}_{\rm g} ) \cdot \textbf{n} \ \mathrm{d}S  \bigg] \nonumber \\
+  \frac{b(\mathbf{x})}{\tau} \bigg[ \int_{V}  \left( \alpha - \alpha_{\mathrm{ref}} \right) \ \mathrm{d}V  \bigg] =  0 
   \quad 
\label{EQtransport_alphaRelax}
\end{align}
with reference velocities $u_{i,\mathrm{ref}}$ and reference volume fraction $\alpha_{\mathrm{ref}}$.
with volume $V $ of control volume (CV)  bounded by the closed surface $\mathrm{S}$, fluid velocity $\mathbf{u}=(u_{1},u_{2},u_{3})^{\mathrm{T}}=(u,v,w)^{\mathrm{T}}$, grid velocity $\textbf{u}_{g} $, unit vector \textbf{n} normal to $S$ and pointing outwards, time $t$, pressure $p$, fluid density $\rho$, components $\tau_{ij}$ of the viscous stress tensor,  unit vector \textbf{i}$_{j}$ in direction $ x_{j} $, volume fraction $ \alpha $ of water, reference velocities $\mathbf{u}_{\mathrm{ref}}$ and reference volume fraction $\alpha_{\mathrm{ref}}$.

\textit{Implicit relaxation zones have three case-dependent parameters:} relaxation parameter $\tau$, blending function $b(\mathbf{x})$, and relaxation zone thickness $x_{\mathrm{d}}$. 

The relaxation parameter $\tau$ has unit $[\mathrm{s}]$ and regulates the magnitude of the source term in such a way that a large value of $\tau$ implicates a small source term and vice versa\footnote{Note that in some publications $\tau$ has been considered a numerical-stability parameter and has therefore occasionally been omitted from Eq. (\ref{EQrelaxprinciple}).  
However, the present work demonstrates the physical meaning of $\tau$ and also that its optimum value does not necessarily coincide with the value which gives the most favorable matrix conditioning.}.  

The blending function $b(\mathbf{x})$ is bounded between $0$ and $1$. In this work, exponential-, cosine- and power-blending functions will be used
\begin{equation}
b(\mathbf{x}) = \left( \frac{e^{( (x_{\mathrm{d}}-\tilde{x}) /x_{\mathrm{d}})^{n}} - 1}{e^{1} - 1} \right) \quad ,
\label{EQblendexp}
\end{equation}
\begin{equation}
b(\mathbf{x}) = \left[\cos^{2}\left( \frac{\pi}{2} + \frac{\pi}{2} \left(\frac{x_{\mathrm{d}}-\tilde{x}}{x_{\mathrm{d}}}\right) \right)\right]^{n} \quad ,
\label{EQblendcos2}
\end{equation}
\begin{equation}
b(\mathbf{x}) = \left( \frac{x_{\mathrm{d}}-\tilde{x}}{x_{\mathrm{d}}} \right)^{n} \quad ,
\label{EQblendpow}
\end{equation}
where $\tilde{x}$ is the shortest distance to the closest domain boundary to which a relaxation zone of thickness $x_{\mathrm{d}}$ is attached (confer Fig. \ref{FIGrelaxdom}), and $n$ regulates the shape of the blending function. Outside the relaxation zone holds $b(\mathbf{x})=0$.

\section{Determining reflection coefficient $C_{\mathrm{R}}$ for implicit relaxation zones}
\label{SECcomputeCR}
For regular, long-crested waves entering a relaxation zone with normal incidence, the reflection coefficient is $C_{\mathrm{R}} = H_{\mathrm{R}}/H$ in terms of the wave heights $ H_{\mathrm{R}} $ and $H$ of the reflected and the generated wave, respectively. 

Following Ursell et al. (1960), $C_{\mathrm{R}}$ can be computed via
\begin{equation}
C_{\mathrm{R}} = \left(H_{\mathrm{max}} - H_{\mathrm{min}}\right)/\left(H_{\mathrm{max}} + H_{\mathrm{min}}\right) \quad , 
\label{EQCRfsursell}
\end{equation}
where $ H_{\mathrm{max}} $ and  $ H_{\mathrm{min}} $ are the overall largest and smallest wave heights that occur, e.g. in this work, during the last simulated wave period over a distance of ca. one wavelength outside but adjacent to the relaxation zone. This approach has a comparatively small background noise of ca. $1\%$ (cf. Perić and Abdel-Maksoud, 2018; Perić, 2019), i.e. reflection coefficients $C_{\mathrm{R}}\lesssim 0.01$ cannot be detected.

However, before the output of Eq. (\ref{EQCRfsursell})  qualifies as reflection coefficient, additional requirements must be fulfilled: The domain size and simulation duration must be chosen so that wave reflections have fully developed in the evaluation interval, while possible wave re-reflections (e.g. at the inlet boundary) have not yet traveled back into the evaluation interval. Further, it must hold  $ 0 \leq C_{\mathrm{R}} \leq 1 $, with $C_{\mathrm{R}} = 0$ for no wave reflection and $C_{\mathrm{R}} = 1$ for perfect wave reflection. Therefore, the boundary conditions must be chosen so that  $C_{\mathrm{R}}=1$ is obtained if the source terms are set to zero, which is discussed in more detail in Sect. \ref{SECfsrelaxrelax2streamfct}. Unless mentioned otherwise, these requirements are fulfilled in the present work.

\section{Analytical approach for optimizing the case-dependent parameters in implicit relaxation zones}
\label{SECtheory}
This section proposes an analytical approach to predict the optimum values for the case-dependent parameters in implicit relaxation zones. For this, the analytical solution from Perić and Abdel-Maksoud (2018) is extended to implicit relaxation zones, be reformulating them into (mathematically) `equivalent forcing zones'.

The implicit relaxation zone can be interpreted as a forcing zone (cf. Perić, 2019), when Eqs. (\ref{EQnavier_stokesRelax}) and (\ref{EQtransport_alphaRelax}) are multiplied by the factor $1/\left( 1-b(\mathbf{x}) \right)$, which gives 
\begin{align}
\frac{\mathrm{d}}{\mathrm{d} t} \int_{V} \rho u_{i} \ \mathrm{d}V 
+ \int_{S} \rho u_{i} (\textbf{u} - \textbf{u}_{g} ) \cdot \textbf{n} \ \mathrm{d}S =  \nonumber \\ 
\int_{S} (\tau_{ij}\textbf{i}_{j} - p\textbf{i}_{i}) \cdot \textbf{n} \ \mathrm{d}S 
+ \int_{V} \rho \textbf{g} \cdot \mathbf{i}_{i} \ \mathrm{d}V + \int_{V} \rho q_{i} \ \mathrm{d}V \quad ,
\label{EQnavier_stokes}
\end{align}
\begin{equation}
\frac{\rm d}{{\rm d} t} \int_{V} \alpha \ \mathrm{d}V + \int_{S} \alpha (\textbf{u} - \textbf{u}_{\rm g} ) \cdot \textbf{n} \ \mathrm{d}S =  \int_V q_{\alpha} \ \mathrm{d}V \quad ,
\label{EQtransport_alpha}
\end{equation}
with forcing source-terms
\begin{align}
q_{i} = \int_{V} \rho \gamma  b(\mathbf{x}) (u_{i,\mathrm{ref}}-u_{i}) \ \mathrm{d}V \quad ,
\label{EQqi}
\end{align}
\begin{equation}
q_{\mathrm{\alpha}} = \int_V \gamma   b(\mathbf{x}) \left( \alpha_{\mathrm{ref}} -\alpha \right) \ \mathrm{d}V \quad ,
\label{EQqa}
\end{equation}
where the forcing strength $\gamma$ is set to 
\begin{equation}
\gamma = \left( \frac{\bar{E}_{\mathrm{kin},x}+\bar{E}_{\mathrm{kin},y}+\bar{E}_{\mathrm{kin},z}+\bar{E}_{\mathrm{pot}}}{\bar{E}_{\mathrm{kin},x}}\right)  \frac{1}{\tau (1-b(\mathbf{x}))} \quad ,
\label{EQrelaxGamma}
\end{equation}
with control volume $V $, fluid density $\rho$, location $\mathbf{x}$, kinetic $\bar{E}_{\mathrm{kin},x_{i}}$ and potential $\bar{E}_{\mathrm{pot}}$ wave energy components\footnote{Note that $\bar{E}_{\mathrm{kin},x_{i}}$ and $\bar{E}_{\mathrm{pot}}$ are not evaluated from the simulation results, but only their ratios are important, which can be taken from wave theory. The reason for the appearance of these terms is that in shallow water, the vertical kinetic energy component vanishes, and consequently the influence of the source terms in the equation for the vertical fluid momentum vanishes. Therefore, to obtain a forcing or relaxation with the same source term magnitude in shallow water as in deep water, the magnitude of the source terms must be increased in shallow water. This effect is comparatively small; it  can change the optimum values for $\gamma$ or $\tau$ by a factor of $2$ at most. Therefore, the factor containing the wave energy components in Eq. (\ref{EQrelaxGamma}) can be computed from linear wave theory.}, relaxation parameter $\tau$ and blending function $b(\mathbf{x})$.

For given blending function $b(\mathbf{x})$ and zone thickness $x_{\mathrm{d}}$, the optimum value of $\gamma$ in Eq. (\ref{EQrelaxGamma}) can be computed analytically as given in Perić and Abdel-Maksoud (2018). Therefore, the optimum value of $\tau$ can be obtained by rearranging Eq. (\ref{EQrelaxGamma}). A simple computer program to optimize implicit relaxation zones has been published as free software: \url{https://github.com/wave-absorbing-layers/relaxation-zones-for-free-surface-waves}.

The derivation of the analytical solution for Eqs. (\ref{EQnavier_stokes}-\ref{EQrelaxGamma}) in Perić and Abdel-Maksoud (2018) neglects some flow phenomena of minor importance, such as that reflected wave components due to source terms in different governing equations can have different phases and may partially cancel destructively. Thus, actual reflection coefficients can be lower than predicted via Eq. (\ref{EQrelaxGamma}). Apart from this, the following can be expected from literature (Perić, 2019; Perić and Abdel-Maksoud, 2016, 2018, 2020):  

The optimum value of relaxation parameter $\tau$ will be closely predicted. The predictions for reflection coefficient $C_{\mathrm{R}}$  can be taken as estimates for the upper-limit of the actual reflection coefficients in the simulations. The implicit relaxation zones behave  discretization-independent for practical discretizations (i.e. more than ca. 30 cells per wavelength).

For irregular waves, the overall reflection coefficient $C_{\mathrm{R}}$ can be estimated based on the reflection coefficients of each wave component.

For nonlinear waves, the analytical approach can be applied without modification,  because, for optimized parameters, partial wave-reflection occurs throughout the relaxation zone with small amplitudes (i.e. nearly linear waves), which can interfere destructively. This ability to `linearize' nonlinear waves makes relaxation zones applicable to highly nonlinear, complex flows, because they produce basically the same amount of reflection regardless of the wave's nonlinearity. This is their main advantage compared to boundary-based approaches such as absorbing boundary conditions, where a complex nonlinear solution must be prescribed at the domain boundary and reflection coefficients can increase unpredictably with flow nonlinearity.

For oblique wave incidence, the analytical approach above can be extended to provide the reflection coefficient as a function of the wave-incidence angle. Results from 3D-flow simulations with strongly reflecting bodies in waves suggest that the analytical approach for 2D-wave propagation as outlined above typically suffices  to optimize the relaxation zone's parameters.

\section{Simulation setup}
\label{SECfsrelaxsetup}
\begin{figure}[h!]
\begin{center}
\includegraphics[width=\linewidth]{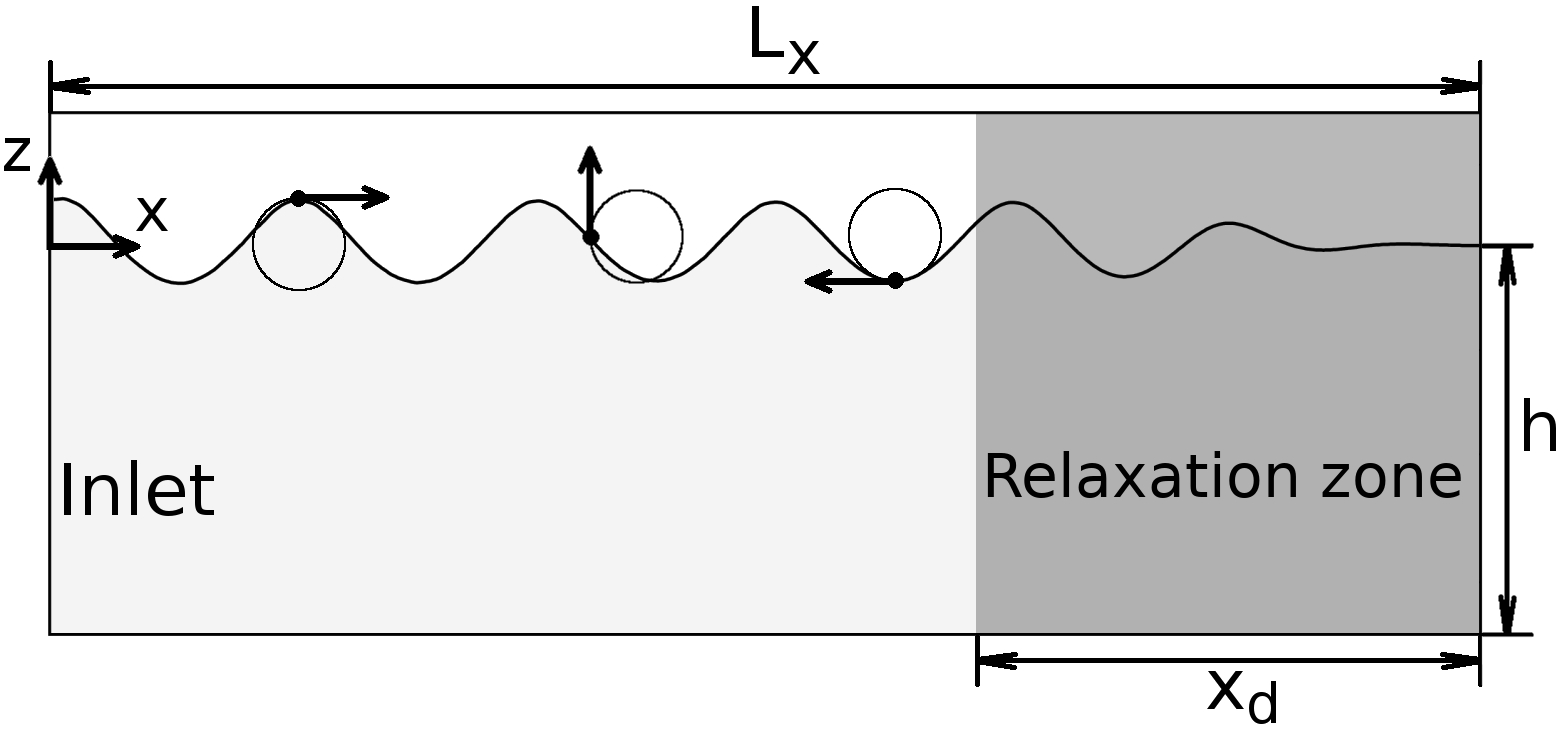}
\end{center}
\caption{Solution domain filled with air (white) and water (light gray, water depth $h$), velocity inlet at $x=0$ and implicit relaxation zone (shaded dark gray) with thickness $x_{\mathrm{d}}$; three fluid particles (black dots) are sketched with their particle paths (circles) and velocity vectors (arrows)} \label{FIGrelaxdom}
\end{figure}

For the 2D-simulations in Sects. \ref{SECfsrelax2Ddeepwaterdampgridstudy} to \ref{SECfsrelaxchoice0blending}, the solution domain is box-shaped as seen in Fig. \ref{FIGrelaxdom}. The origin of the coordinate system lies at the calm free-surface level, with $z$ pointing upwards and $x$ pointing in wave propagation direction. The domain dimensions are $ 0\, \mathrm{m}\leq x \leq 24\, \mathrm{m}$,  $-2\, \mathrm{m}\leq z \leq 0.24\, \mathrm{m}$ for the simulations with deep water conditions (water depth $h \approx 0.5 \lambda$) and $ 0\, \mathrm{m}\leq x \leq 24\, \mathrm{m}$,  $-0.2\, \mathrm{m}\leq z \leq 0.01\, \mathrm{m}$ for the simulations with shallow water conditions ($h \approx 0.05 \lambda$).
The simulations are quasi-2D, i.e. there is only one layer of cells in $y$-direction and the $y$-normal boundaries set to symmetry planes.

Waves are generated by prescribing volume fraction $\alpha $ and velocities $\mathbf{u}$ according to Rienecker and Fenton's (1981) stream function wave theory ($64^{\mathrm{th}}$ order) at the velocity inlet $x=0$. Table \ref{TABwaveParameters} gives the wave parameters.
The waves travel in positive $x$-direction towards an implicit  relaxation zone attached to the pressure outlet boundary at $x=24\, \mathrm{m}$. At the outlet, pressure and volume fraction are prescribed according to the calm free-surface solution. 

\begin{table}
\caption{Wave parameters for the different simulation setups: wave height $H$, wave period $T$, wavelength $\lambda$, water depth $h$, wave steepness $H/\lambda$ in terms of maximum wave steepness $(H/\lambda)_{\mathrm{max}}$}
\label{TABwaveParameters}
{\renewcommand{\arraystretch}{1.2}%
\begin{tabular}{rccccc}
 & $H$ & $T$ & $\lambda$  & $\frac{H/\lambda}{(H/\lambda)_{\mathrm{max}}}$  \vspace*{0.08cm}\\ 
\hline
2D, deep water & $ 0.16\, \mathrm{m} $ & $1.6\, \mathrm{s}$ & $ 4\, \mathrm{m} $  & $ 28.6\% $\\ 
2D, shallow water & $0.009\, \mathrm{m}$ & $ 2.893\, \mathrm{s} $ & $ 4\, \mathrm{m} $&  $22.5\%$\\ 
3D, deep water & $ 0.4\, \mathrm{m} $ & $ 1.6\, \mathrm{s} $ & $4.3\, \mathrm{m}$ & $65.5\%$
\end{tabular} 
}
\end{table}

The governing equations are Eqs. (\ref{EQnavier_stokesRelax}) to (\ref{EQtransport_alphaRelax}). In the implicit relaxation zone, velocity $\mathbf{u}$ and volume fraction $\alpha$ are blended towards the analytical reference solution, $\mathbf{u}_{\mathrm{ref}}$ and $\alpha_{\mathrm{ref}}$, to reduce undesired wave reflections. The relaxation zone's parameters are optimized according to the analytical approach presented in Sect. \ref{SECtheory}. Simulations are performed for different values of zone thickness $x_{\mathrm{d}}$, different blending functions $b(\mathbf{x})$, and different reference solutions. The bottom boundary  has a slip-wall boundary condition and at the top boundary atmospheric pressure is prescribed. 

The simulations in this work are performed using  \texttt{foam-extend} version 4.1, an open-source fork of the flow-solver OpenFOAM (Weller et al., 1998),  combined with the commercial software Naval Hydro Pack. 
The governing equations are Eqs. (\ref{EQnavier_stokesRelax}) to (\ref{EQtransport_alphaRelax}), so no turbulence modeling is used.  All approximations are of second order.
The solvers are conjugate gradient with Incomplete Cholesky preconditioner for pressures and bi-conjugate gradient with ILU0 preconditioner for volume fraction and velocities. The PIMPLE scheme is used with two pressure-correction steps per each of the two nonlinear iterations per time step. No under-relaxation is used. In all simulations, the Courant number $C=|\mathbf{u}|\Delta t / \Delta x$ remains well below $0.4$.  Further information on the discretization of and solvers for the governing equations can be found in  Ferziger and Peri\'c (2020) and the flow solver manuals. 

\begin{figure}[h!]
\begin{center}
\includegraphics[width=\linewidth]{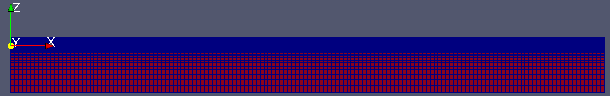}\\
\includegraphics[width=\linewidth]{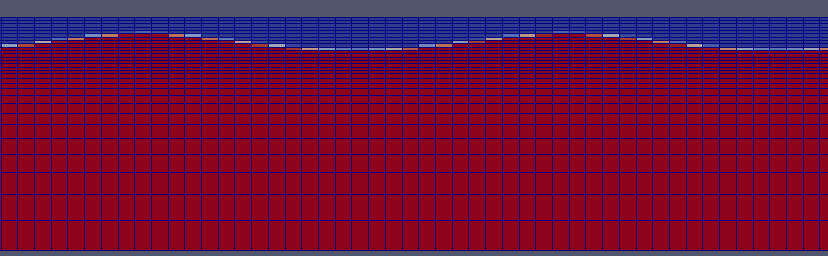}
\end{center}
\caption{Mesh for 2D-flow simulations with coarse grid; far view (top) and close-up (bottom); the color denotes the volume fraction (red: water, blue: air) } \label{FIG2Dmesh}
\end{figure}

Figure \ref{FIG2Dmesh} shows the rectilinear grid  with local mesh refinement. The free surface remains at all times within the zone with the finest mesh, with $ 25$ (coarse grid), $ 35 $ (medium grid), or $ 50 $ (fine grid) cells per wavelength $\lambda$, and $ 5 $ (coarse grid), $ 7 $ (medium grid), or $ 10 $ (fine grid) cells per wave height $H$. The grid consists of $ 12\, 000 $ (coarse grid), $ 27\, 000 $ (medium grid), or $ 48\, 000 $ (fine grid) cells. 
The time-step is $ 0.01\, \mathrm{s} = T/160 $ (coarse grid), $ 0.0071\, \mathrm{s} = T/226 $ (medium grid), or $ 0.005\, \mathrm{s} = T/320 $ (fine grid). The reflection coefficient $C_{\mathrm{R}}$ is calculated as given in Sect. \ref{SECcomputeCR}.

In Sect. \ref{SECfsrelax2Dstarfoamcomp}, selected simulations were repeated with the commercial flow solver STAR-CCM+ version 10.6 by Siemens, using the grid and simulation setup from Perić and Abdel-Maksoud (2018), except that the forcing zones in STAR-CCM+ were optimized via Eqs. (\ref{EQqi}) to (\ref{EQrelaxGamma}) in such a way that they mimic the behavior of the implicit relaxation zones from the Naval Hydro Pack.

\begin{figure}[h!]
\begin{center}
\includegraphics[width=\linewidth]{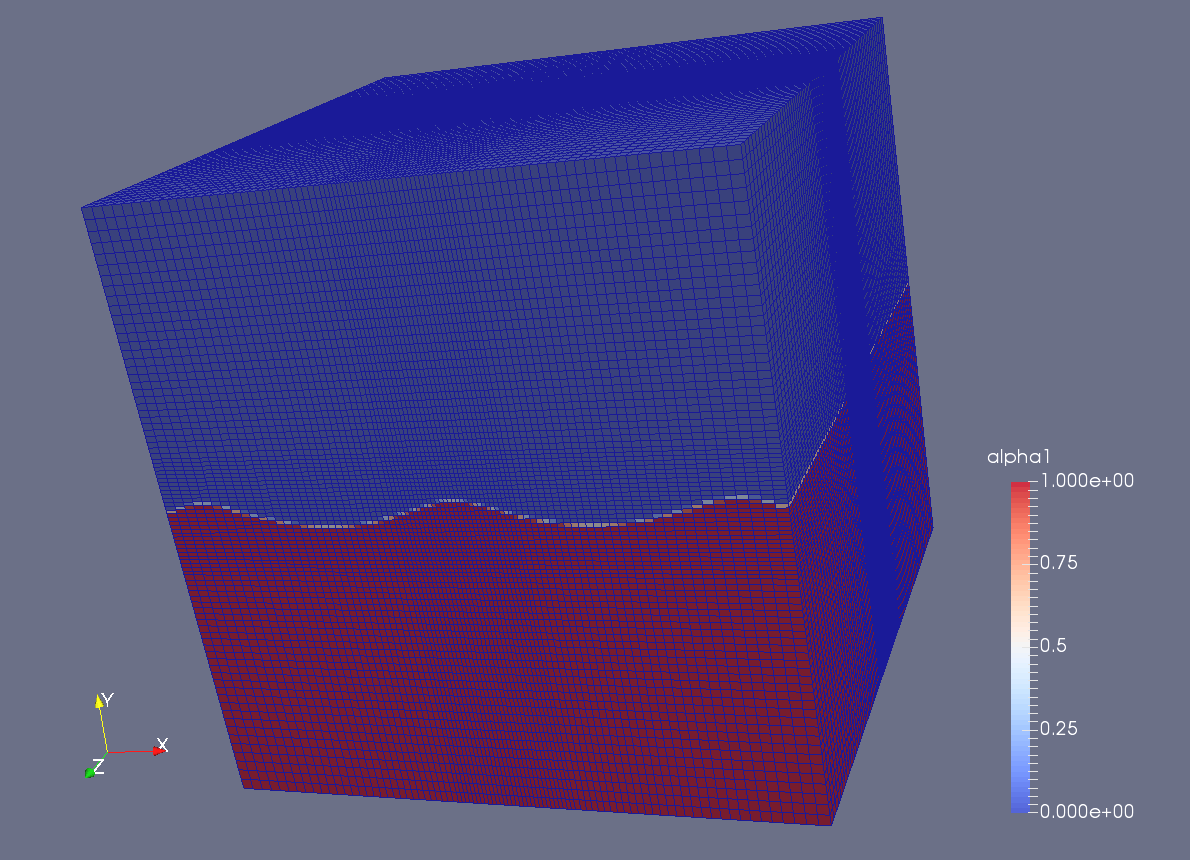}
\end{center}
\caption{Fine mesh for 3D-flow simulations with initialized volume fraction} \label{FIG3Dmesh}
\end{figure}

For the 3D-simulations in Sect. \ref{SECfsrelaxres3d}, the setup is identical to the deep-water 2D simulations, with the following exceptions. The domain has dimensions $ 0\, \mathrm{m}\leq x \leq 10\, \mathrm{m}$, $0\, \mathrm{m}\leq y \leq 10\, \mathrm{m}$, $-5\, \mathrm{m}\leq z \leq 5\, \mathrm{m}$, so the water depth is $h=5\, \mathrm{m}$ as seen in Fig. \ref{FIG3Dmesh}. In the center of the domain, a semi-submerged pontoon with dimensions $1\, \mathrm{m} \times1\, \mathrm{m} \times1\, \mathrm{m}$ is held in fixed position as seen in Fig. \ref{FIG3Dfsfromvids}. It has a draft of $D=0.5\, \mathrm{m}$ and slip wall boundary conditions. The wave parameters are given in Table \ref{TABwaveParameters}.

The relaxation zone thickness is $x_{\mathrm{d}}=3\, \mathrm{m}\approx 0.7\lambda$ and power blending according to Eq. (\ref{EQblendpow}) with exponent $n=0.46$ is used. Simulations are performed for different relaxation parameters $0.001\, \mathrm{s}\leq \tau \leq 1000\, \mathrm{s}$. This setup is expected to be close to the minimum domain size for the simulation of such a strongly wave-reflecting body.

The free surface is discretized by  $12.9$ (coarse grid), $25.8$ (medium grid), or $38.7$ (fine grid) cells per wavelength $\lambda$ and $2$ (coarse grid), $4$ (medium grid), or $6$ (fine grid) cells per wave height $H$ as shown in Fig. \ref{FIG3Dmesh}. Per wave period $160$ (coarse grid), $225$ (medium grid), or $320$ (fine grid) time steps are used.

\section{Results from 2D-flow simulations}
\label{SEC2dflow}
This section compares the analytical predictions from Sect. \ref{SECtheory} against results from 2D-flow simulations based on the 2D-setup from Sect. \ref{SECfsrelaxsetup}.

\subsection{Discretization dependence study for wave damping via implicit relaxation zones in deep  water}
\label{SECfsrelax2Ddeepwaterdampgridstudy}
This section investigates wave damping via implicit relaxation zones in deep water. To damp the waves, the reference solution in Eqs. (\ref{EQnavier_stokesRelax}) and (\ref{EQtransport_alphaRelax}) is set to the hydrostatic solution for the calm free-surface. Exponential blending  via Eq. (\ref{EQblendexp}) with coefficient $n=3.5$ is used, which is the default setting in the Naval Hydro Pack. Simulations are performed for different values of zone thickness $x_{\mathrm{d}}$ and relaxation parameter $\tau$. 

\begin{figure}[h!]
\begin{center}
\includegraphics[width=\linewidth]{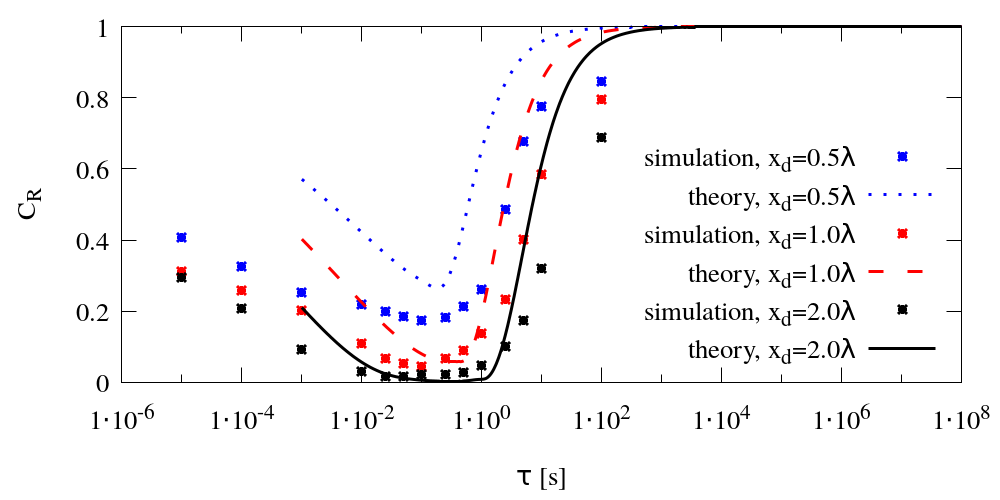}
\includegraphics[width=\linewidth]{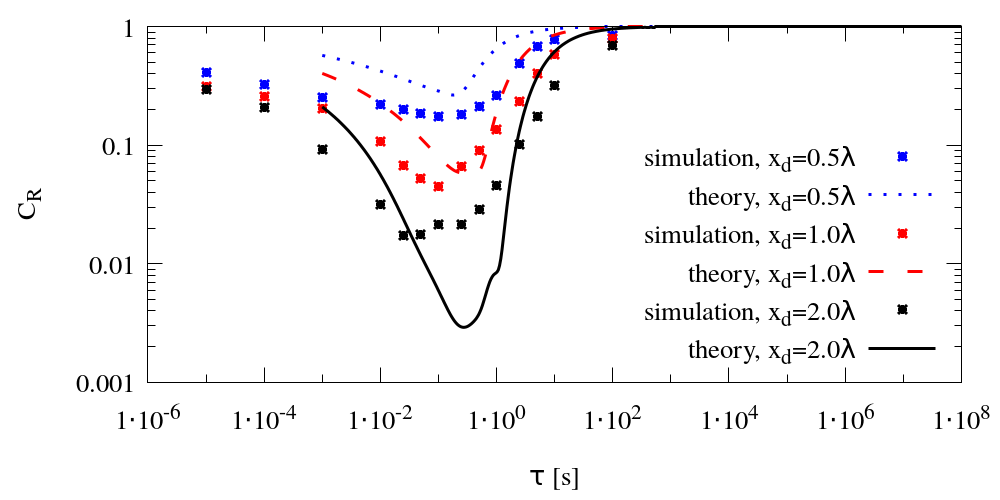}
\end{center}
\caption{Analytical predictions (`theory') and simulation results (`simulation') for reflection coefficient $C_{\mathrm{R}}$ as a function of relaxation parameter $\tau$, for deep-water waves with period $T=1.6\, \mathrm{s}$; for exponential blending via Eq. (\ref{EQblendexp}) with exponent $n=3.5$, coarse discretization and different values of relaxation zone thickness $x_{\mathrm{d}}$; for all simulation results $C_{\mathrm{R,sim}}$ and corresponding analytical predictions $C_{\mathrm{R,theory}}$ holds $C_{\mathrm{R,sim}} - C_{\mathrm{R,theory}} < 3.8\%$; for the forcing strength $\tau \leq \tau_{\mathrm{opt,theory}}$ closest to the theoretical optimum value $\tau_{\mathrm{opt,theory}}$ holds $C_{\mathrm{R,sim}} - C_{\mathrm{R,theory}} < 1.9\%$} \label{FIG2Dexpn3p5coarseDeep}
\end{figure}

Figure \ref{FIG2Dexpn3p5coarseDeep} demonstrates  for different zone thicknesses $x_{\mathrm{d}}$ that the analytical approach proposed in Sect. \ref{SECtheory} predicts the optimum value of relaxation parameter $\tau$ closely. 
As expected from the discussion in Sect. \ref{SECtheory}, the analytical predictions  additionally provide a satisfactory estimate of the upper-bound for reflection coefficient $C_{\mathrm{R}}$.

Note that the lower plot in Figs. \ref{FIG2Dexpn3p5coarseDeep}, \ref{FIG2Dexp3p5coarseShallow} and \ref{FIG2Dexpn3p5coarseDeepRelax2Streamfct} shows the same data as the upper plot but with a logarithmic vertical axis, to better visualize the results for small values of reflection coefficient $C_{\mathrm{R}}$. Furthermore, note that due to the slight background-noise in the scheme for determining $C_{\mathrm{R}}$, reflection coefficients below ca. $0.01$ cannot be detected reliably. Finally, note that the curves in each plot hold only for the given wave period $T$; when the wave period changes, the curves shift sideways.

\begin{figure}[h!]
\begin{center}
\begin{small}
coarse discretization\\
\includegraphics[width=\linewidth]{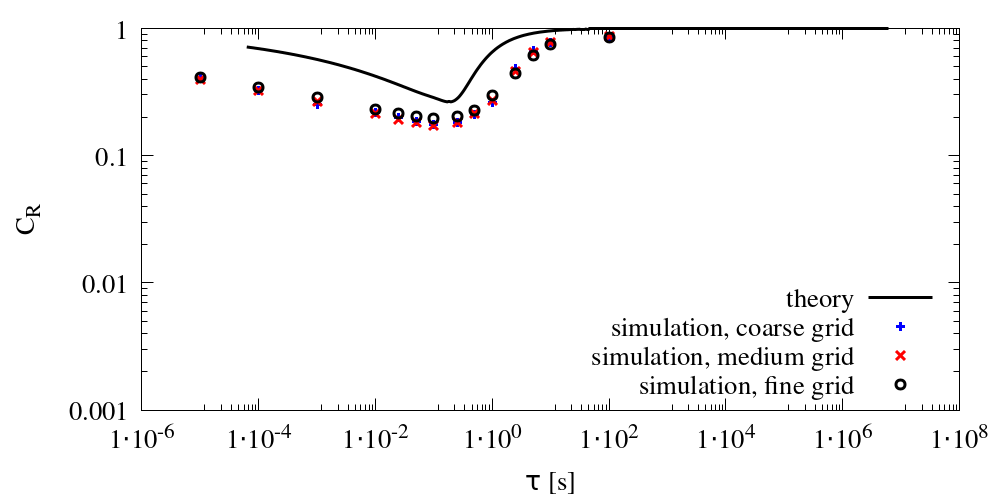} \\
medium discretization \\ 
 \includegraphics[width=\linewidth]{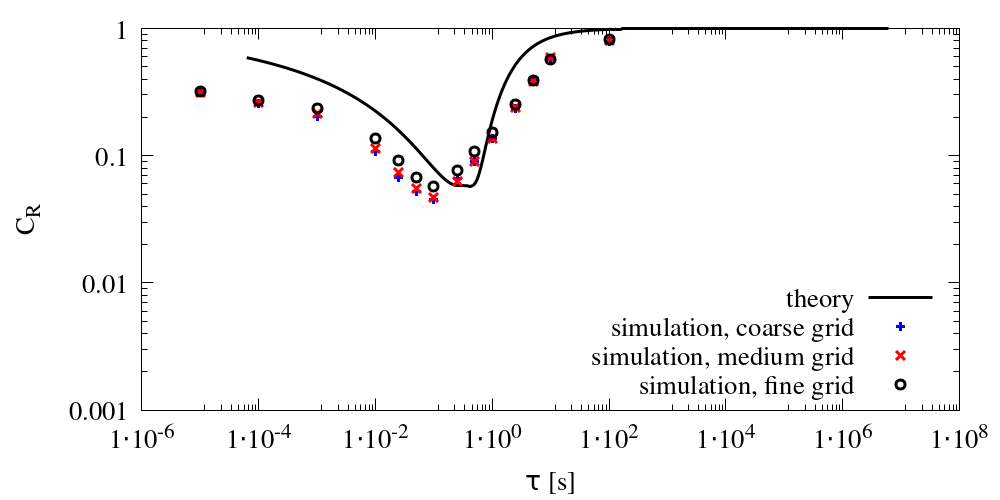} \\  
fine discretization   \\ 
\includegraphics[width=\linewidth]{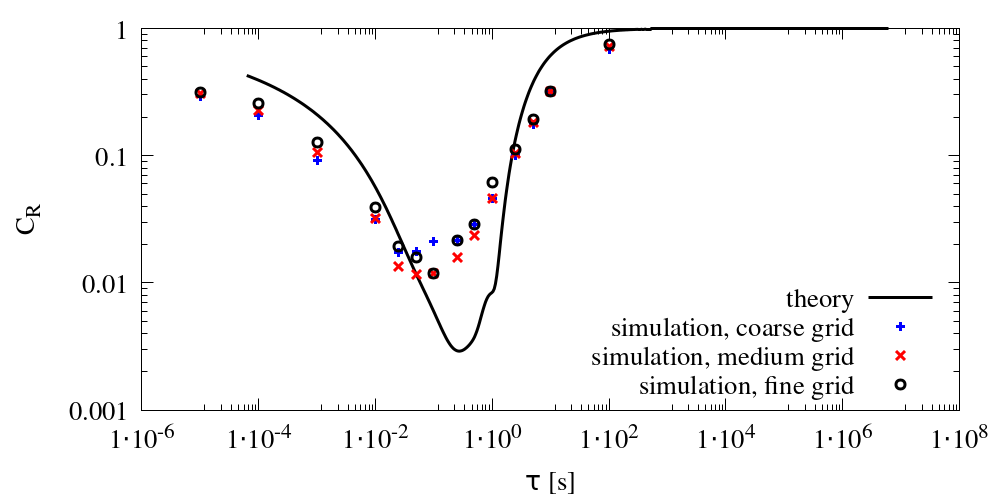}  
\end{small}
\end{center}
\caption{As Fig. \ref{FIG2Dexpn3p5coarseDeep}, except for coarse, medium, and fine discretization from Sect. \ref{SECfsrelaxsetup}; as theory suggests, results can be considered discretization-independent for practical discretizations; for all simulation results $C_{\mathrm{R,sim}}$ and corresponding analytical predictions $C_{\mathrm{R,theory}}$ holds $C_{\mathrm{R,sim}} - C_{\mathrm{R,theory}} < 5.3\%$; for the relaxation parameter $\tau \leq \tau_{\mathrm{opt,theory}}$ closest to the theoretical optimum value $\tau_{\mathrm{opt,theory}}$ holds $C_{\mathrm{R,sim}} - C_{\mathrm{R,theory}} < 1.9\%$} \label{FIG2Dexpn3p5cmfDeep}
\end{figure}

Figure \ref{FIG2Dexpn3p5cmfDeep} demonstrates that, for practical discretizations (i.e. $25$ or more cells per wavelength $\lambda$), implicit relaxation zones behave  basically discretization-independent. The slight background-noise in the scheme for determining reflection coefficient $C_{\mathrm{R}}$ was attributed to the interface-sharpening scheme (cf. Larsen et al., 2019; Berndt et al., 2021).

\subsection{Wave damping via implicit relaxation zones in shallow water}
\label{SECfsrelax2Dshallowwaterdamp}
\begin{figure}[h!]
\begin{center}
\includegraphics[width=\linewidth]{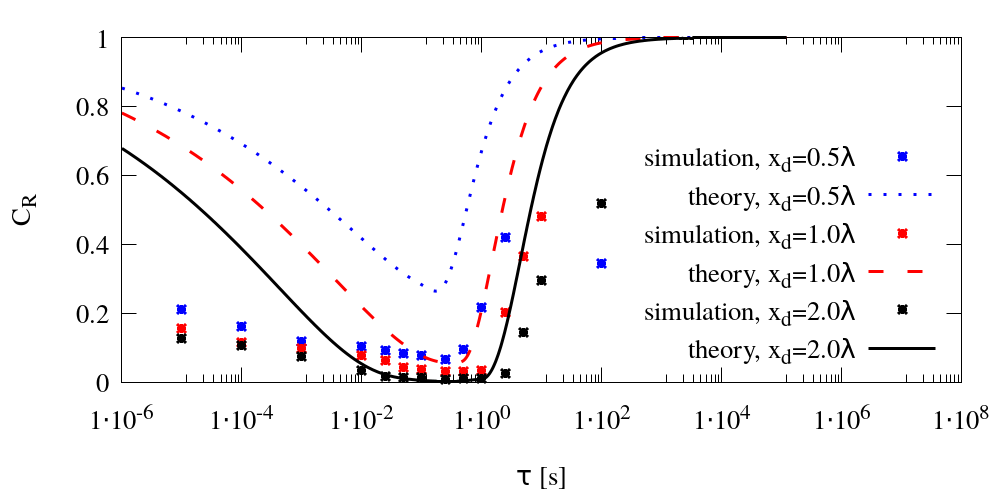}
\includegraphics[width=\linewidth]{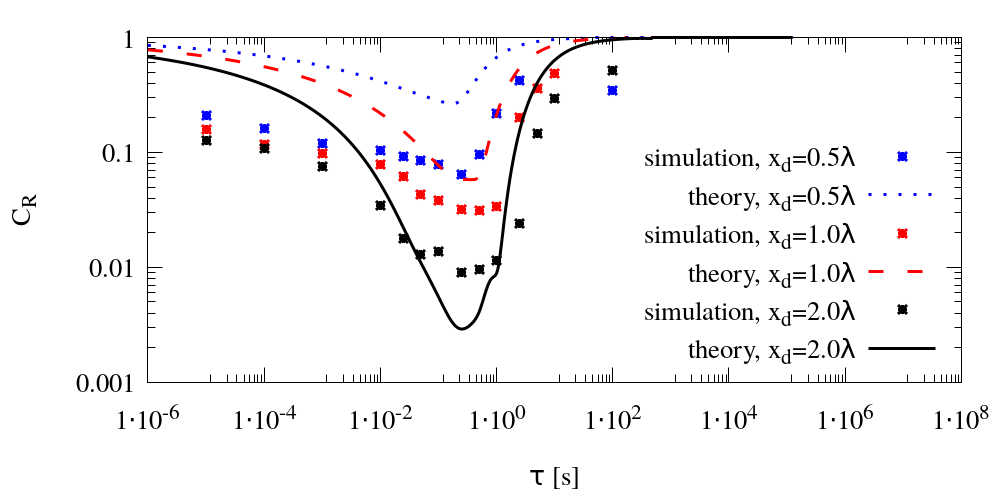}
\end{center}
\caption{As Fig. \ref{FIG2Dexpn3p5coarseDeep}, except for shallow-water waves with period $T=2.893\, \mathrm{s}$; for all simulation results $C_{\mathrm{R,sim}}$ and corresponding analytical predictions $C_{\mathrm{R,theory}}$ holds $C_{\mathrm{R,sim}} - C_{\mathrm{R,theory}} < 0.8\%$; for the forcing strength $\tau \leq \tau_{\mathrm{opt,theory}}$ closest to the theoretical optimum value $\tau_{\mathrm{opt,theory}}$ holds $C_{\mathrm{R,sim}} - C_{\mathrm{R,theory}} < 0.5\%$} \label{FIG2Dexp3p5coarseShallow}
\end{figure}

This section investigates wave damping via implicit relaxation zones in shallow water. 
The derivation in Sect. \ref{SECtheory} holds for all water depths, and Fig. \ref{FIG2Dexp3p5coarseShallow} confirms that its predictions are of satisfactory accuracy also in shallow water. 

Compared to the deep-water case from Sect. \ref{SECfsrelax2Ddeepwaterdampgridstudy}, the simulation results for  reflection coefficient $C_{\mathrm{R}}$ are lower for smaller-than-optimum values of relaxation parameter $\tau$, but show no substantial qualitative difference otherwise. This was expected, because Perić (2019) showed that in shallow water (where the horizontal components of the average kinetic wave energy are much larger than the vertical component), stronger-than-optimum forcing of volume fraction $\alpha$ reflects waves with a phase shift of $180\, \mathrm{deg}$  compared to forcing of horizontal velocity $u$, so that combined $\alpha$- and $u$-forcing produces destructive interference and thus lower reflection coefficients $C_{\mathrm{R}}$ than in deep water (where the  horizontal and vertical components of the average kinetic energy have the same magnitude).

Recently, Carmigniani and Violeau (2018) used forcing zones for horizontal and vertical velocities to damp regular waves in finite-difference-based flow simulations for linearized Navier-Stokes-equations; they observed a decrease in the optimum value of the source term strength for decreasing water depth. In contrast, the present results show no significant dependence of the optimum value of relaxation parameter $\tau$ on the water depth. 

However, one should point out that, in Figs. \ref{FIG2Dexpn3p5coarseDeep} to \ref{FIG2DpowerNcoarseDeep},  the optimum $\tau$-value from the simulation results is sometimes slightly larger or smaller than predicted analytically. 
The relaxation parameter $\tau_{\mathrm{opt,sim}}$ for the simulation result with the lowest reflection coefficient $C_{\mathrm{R}}$  took values within $\tau_{\mathrm{opt,sim}} \in [\frac{1}{12}\tau_{\mathrm{opt,theory}},2\tau_{\mathrm{opt,theory}}]$, where  $\tau_{\mathrm{opt,theory}}$ denotes  the theoretically predicted optimum $\tau$-value. 
Since there did not seem to be a clear trend in these deviations and since they were comparatively small, this detail seems to be of minor importance for engineering practice.

\subsection{Comparison between implicit relaxation zones and equivalent forcing zones in a different flow solver}
\label{SECfsrelax2Dstarfoamcomp}
This section aims to validate the finding from Sect. \ref{SECtheory}, that implicit relaxation zones can be interpreted as a special-case of forcing zones. For this, flow simulations with wave damping via implicit relaxation zones are performed for different blending functions $b(\mathbf{x})$ with a similar setup as in Sect. \ref{SECfsrelax2Ddeepwaterdampgridstudy}. Then, the flow simulations are repeated using a different flow solver, Siemens STAR-CCM+, with an `equivalent forcing zone' instead of the implicit relaxation zone, and the results are compared.  

The `equivalent forcing zone' is constructed as follows: In STAR-CCM+, forcing zones according to Eqs. (\ref{EQnavier_stokes}) and (\ref{EQtransport_alpha}) are available. 
To `mimick' the behavior of an implicit relaxation zone,  the forcing strength $\gamma$ is selected as given in Eq.  (\ref{EQrelaxGamma}).

\begin{figure}[h!]
\begin{center}
\begin{small}
\texttt{foam-extend} Naval Hydro Pack \\
\includegraphics[width=\linewidth]{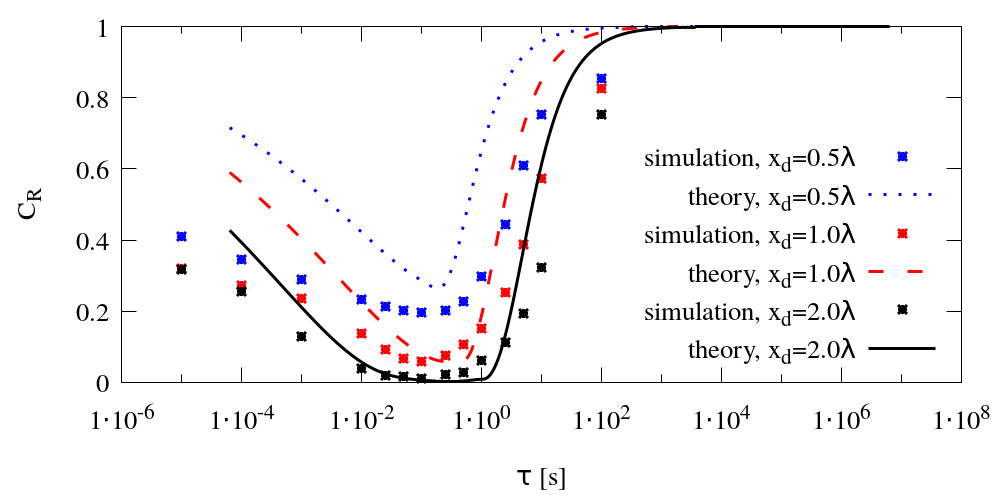} \\
Siemens STAR-CCM+ \\ 
 \includegraphics[width=\linewidth]{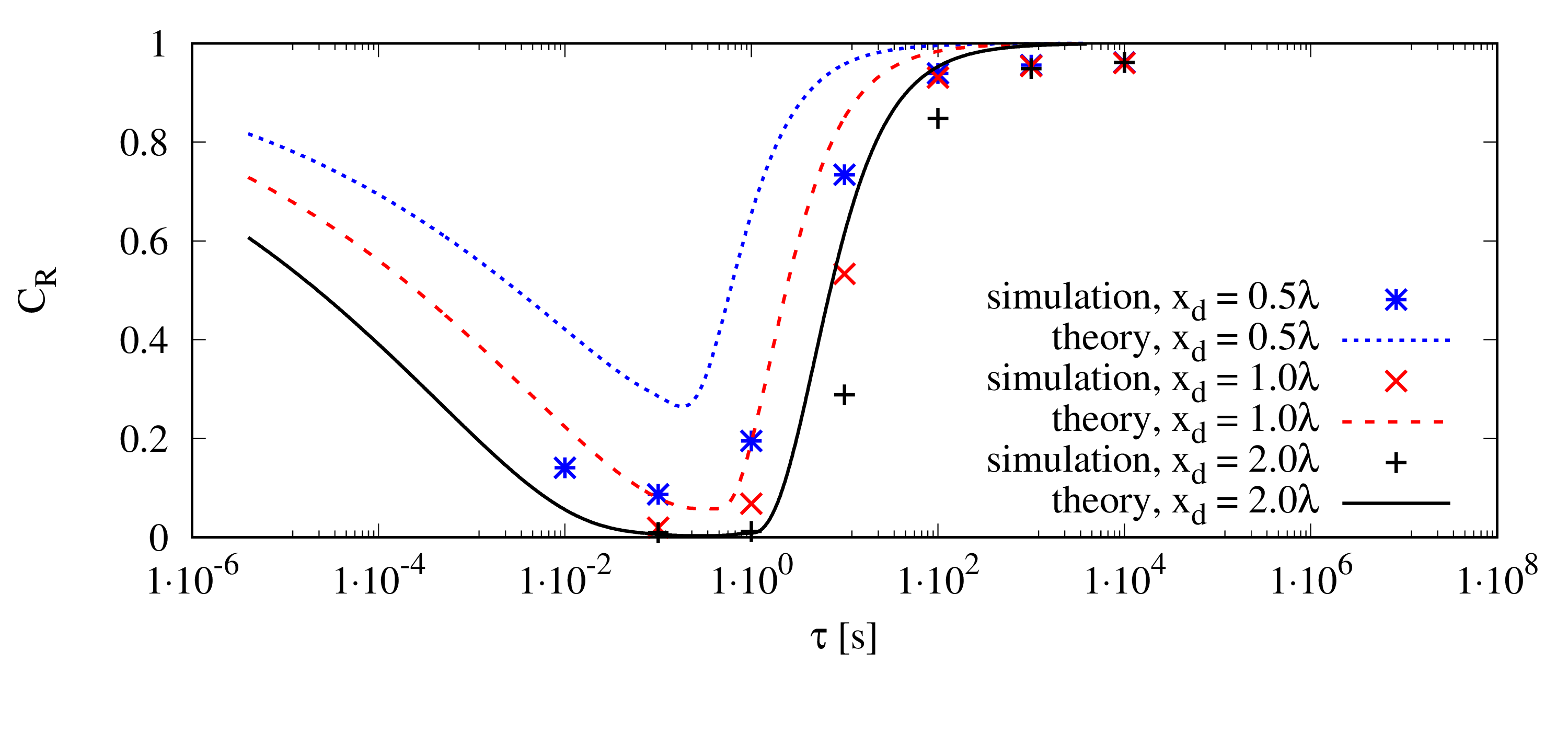} \\ 
\end{small}
\end{center}
\caption{Analytical predictions (`theory') and simulation results (`simulation') from two different CFD codes for reflection coefficient $C_{\mathrm{R}}$ as a function of relaxation parameter $\tau$, for deep-water waves with period $T=1.6\, \mathrm{s}$; for fine discretization, exponential blending via Eq. (\ref{EQblendexp}) with exponent $n=3.5$ and different values of zone thickness $x_{\mathrm{d}}$; for implicit relaxation zone (top) and its `equivalent forcing zone' (bottom)} \label{FIG2Dexpn3p5DeepCCMvsFOAM}
\end{figure}

\begin{figure}[h!]
\begin{center}
\begin{small}
\texttt{foam-extend} Naval Hydro Pack \\
\includegraphics[width=\linewidth]{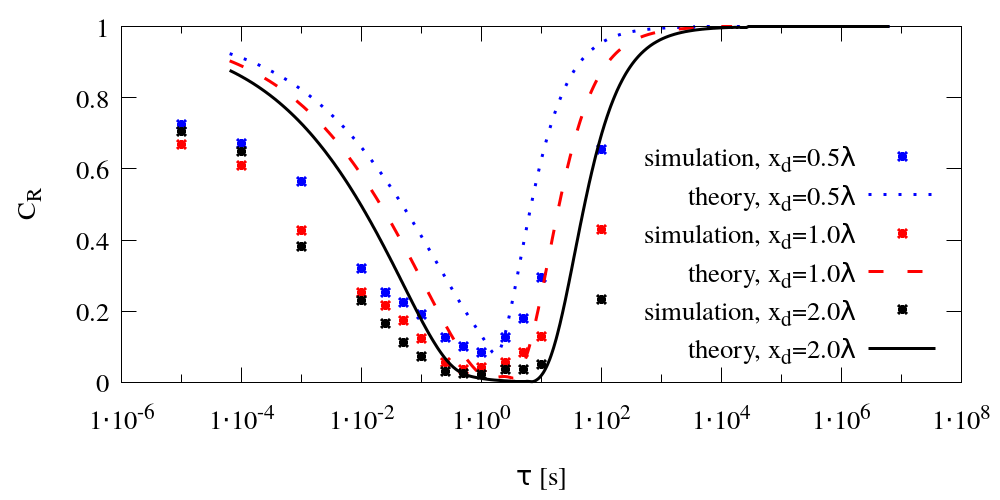} \\
Siemens STAR-CCM+ \\ 
 \includegraphics[width=\linewidth]{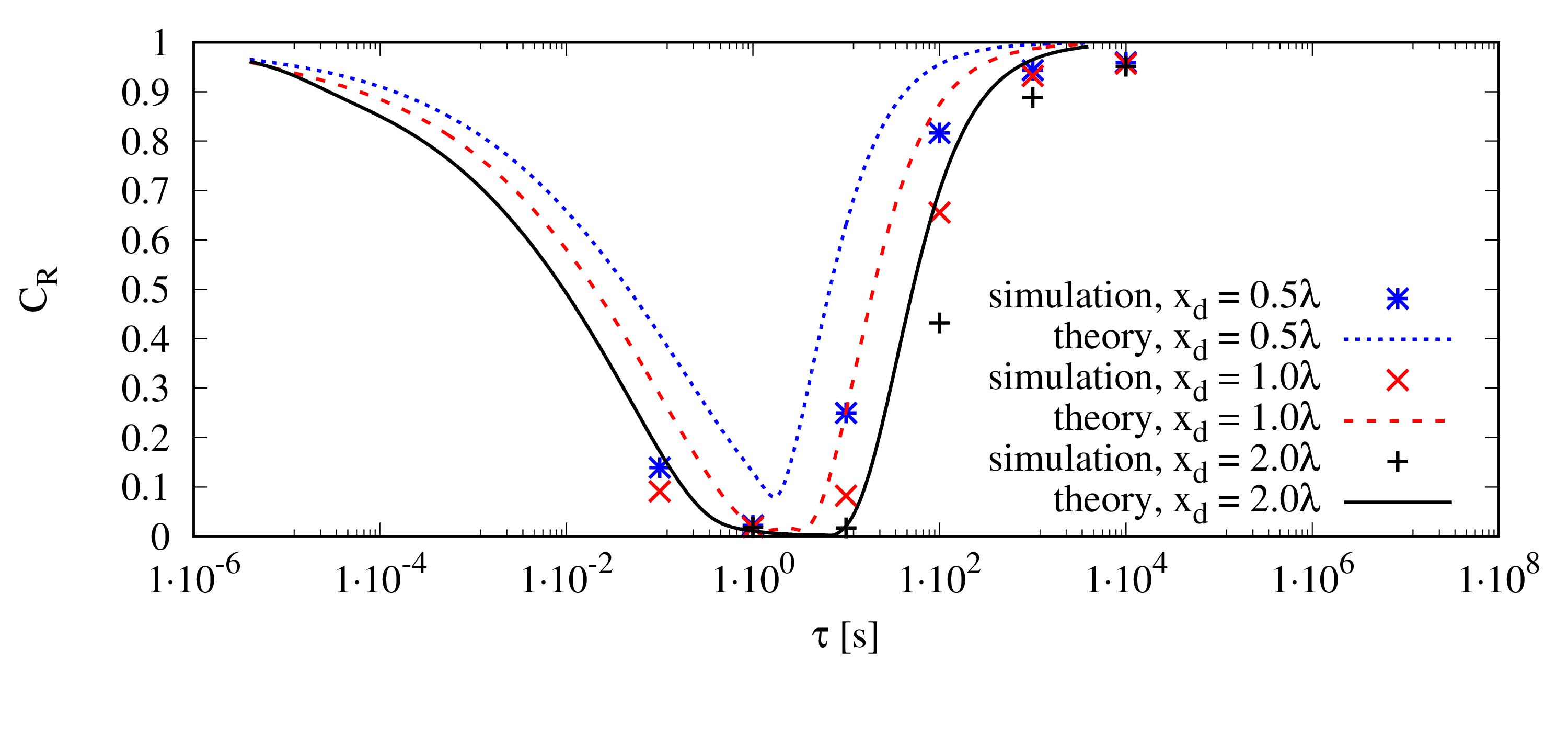} \\ 
\end{small}
\end{center}
\caption{As Fig. \ref{FIG2Dexpn3p5DeepCCMvsFOAM}, except that linear blending according to Eq. (\ref{EQblendpow}) with $n=1$ was used; for all simulation results $C_{\mathrm{R,sim}}$ and corresponding analytical predictions $C_{\mathrm{R,theory}}$ holds $C_{\mathrm{R,sim}} - C_{\mathrm{R,theory}} < 3.9\%$; for the forcing strength $\tau \leq \tau_{\mathrm{opt,theory}}$ closest to the theoretical optimum value $\tau_{\mathrm{opt,theory}}$ holds $C_{\mathrm{R,sim}} - C_{\mathrm{R,theory}} < 3.4\%$} \label{FIG2DlinDeepCCMvsFOAM}
\end{figure}

Note that `mimicking' relaxation zones via forcing zones is performed here only to demonstrate the close relationship between both approaches. In practice, such `mimicking' is not recommended: Even if it results should theoretically be the same, `mimicking' can impair the numerical stability. For example, in the `equivalent forcing zone' outlined above  holds near the domain boundary $b(\mathbf{x}) \rightarrow1 $, so that the source term magnitude would approach infinity. Thus stability problems must be expected for the `equivalent forcing zone' when relaxation parameter $\tau\rightarrow 0$ and when the cell sizes close to the domain boundary are small. 
The STAR-CCM+ simulations indeed blew up for small $\tau$-values, which is the reason for the missing data points ($\tau \leq 10^{-2}\, \mathrm{s}$) in Figs. \ref{FIG2Dexpn3p5DeepCCMvsFOAM} and \ref{FIG2DlinDeepCCMvsFOAM}. 

No stability issues  occur when forcing zones (Eqs. (\ref{EQnavier_stokes}) to (\ref{EQtransport_alpha})) or implicit relaxation zones (Eqs. (\ref{EQnavier_stokesRelax}) and (\ref{EQtransport_alphaRelax})) are used in the way they were intended, as the results in the other sections or in literature (e.g. Perić, 2019; Perić and Abdel-Maksoud, 2018, 2020)  demonstrate.
Further, comparing the present results to the ones from literature indicates that forcing zones and implicit relaxation zones both work equally satisfactory when correctly set up.

Figures \ref{FIG2Dexpn3p5DeepCCMvsFOAM} and \ref{FIG2DlinDeepCCMvsFOAM} show that the results of the two different codes agree well. Thus one can confidently expect both the present results and the analytical approach from Sect. \ref{SECtheory} to be applicable to other computational-fluid-dynamics solvers as well. 

\subsection{Relaxation towards far-field wave vs. relaxation towards calm-water solution}
\label{SECfsrelaxrelax2streamfct}
This section investigates the influence of the choice of  reference solution for implicit relaxation zones. 
In practice,   the reference solution  is often the far-field wave solution. Therefore, the simulations from Sect. \ref{SECfsrelax2Ddeepwaterdampgridstudy} were repeated with   reference solution $\mathbf{u}_{\mathrm{ref}}$ and  $\alpha_{\mathrm{ref}}$ set to the stream function solution for the far-field wave. 

\begin{figure}[h!]
\begin{center}
\begin{small}
coarse discretization \\
\includegraphics[width=\linewidth]{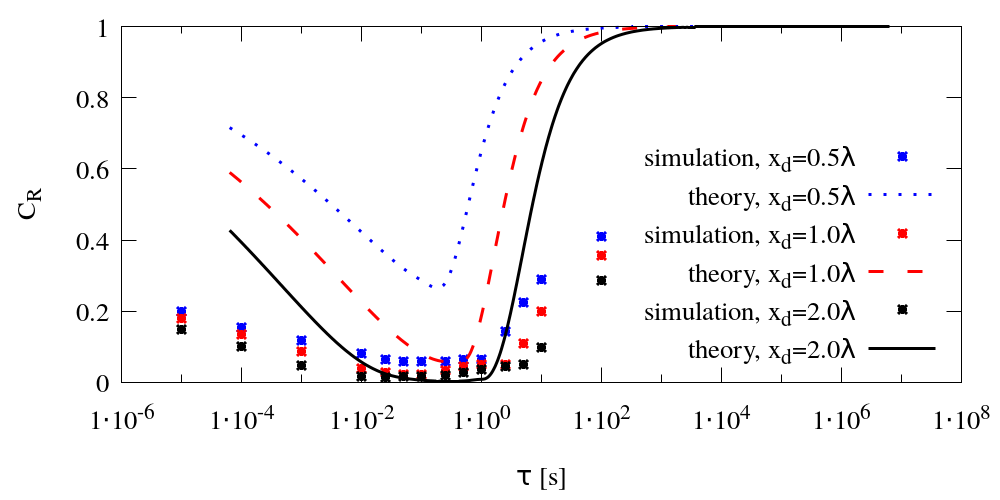} \\
fine discretization \\ 
 \includegraphics[width=\linewidth]{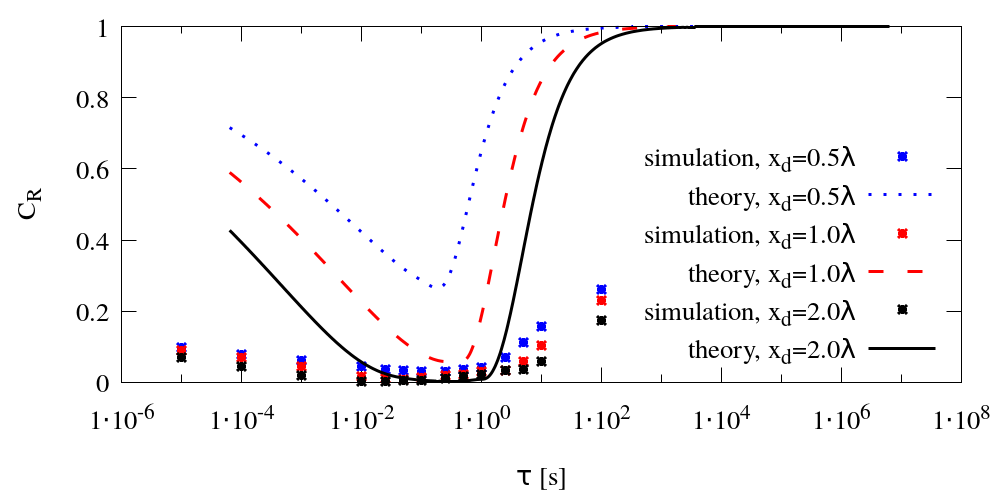} \\ 
\end{small}
\end{center}
\caption{As Fig. \ref{FIG2Dexpn3p5coarseDeep}, except for relaxation towards the far-field wave; for all simulation results $C_{\mathrm{R,sim}}$ and corresponding analytical predictions $C_{\mathrm{R,theory}}$ holds $C_{\mathrm{R,sim}} - C_{\mathrm{R,theory}} < 2.9\%$; for the forcing strength $\tau \leq \tau_{\mathrm{opt,theory}}$ closest to the theoretical optimum value $\tau_{\mathrm{opt,theory}}$ holds $C_{\mathrm{R,sim}} - C_{\mathrm{R,theory}} < 1.6\%$} \label{FIG2Dexpn3p5coarseDeepRelax2Streamfct}
\end{figure}

Figure \ref{FIG2Dexpn3p5coarseDeepRelax2Streamfct} shows that, although Sects. \ref{SECfsrelax2Ddeepwaterdampgridstudy} and \ref{SECfsrelaxrelax2streamfct} use substantially different reference solutions, again the optimum value for relaxation parameter $\tau$ is well predicted. 

However, compared to Fig. \ref{FIG2Dexpn3p5coarseDeep}, the values for reflection coefficient $C_{\mathrm{R}}$ in Fig. \ref{FIG2Dexpn3p5coarseDeepRelax2Streamfct} are substantially lower, which becomes more pronounced on the fine discretization.
The reason for this is that the $C_{\mathrm{R}}$-values in Fig. \ref{FIG2Dexpn3p5coarseDeepRelax2Streamfct} do not qualify as reflection coefficients (cf. definition in Sect. \ref{SECcomputeCR}), because the simulation setup does not not yield $C_{\mathrm{R}} = 1$ if the relaxation zone is switched off: There is no flow disturbing body within the domain, so differences between the computed and reference solution are mainly due to discretization and iteration errors, which vanish on the finer grids;  the smaller the differences between computed and reference solution are, the smaller will be wave reflections at the outlet boundary, where the reference solution is prescribed.
In practice, the wave entering the relaxation zone usually does not correspond to the far-field wave, because it will be modified by wave reflecting bodies or discretization and iteration errors within the domain. 
Thus, for the general case of relaxation towards the far-field wave, one should rather expect reflection coefficients $C_{\mathrm{R}}$ as in Fig. \ref{FIG2Dexpn3p5coarseDeep}.

\subsection{Influence of choice of blending function $b(\mathbf{x})$}
\label{SECfsrelaxchoice0blending}
This section investigates how changing the blending function $b(\mathbf{x})$ can affect the behavior of implicit relaxation zones. For this, the simulations from Sect. \ref{SECfsrelax2Ddeepwaterdampgridstudy} were repeated using different blending functions. 

\begin{figure}[h!]
\begin{small}
\begin{center}
$n = 10$\\
\includegraphics[width=\linewidth]{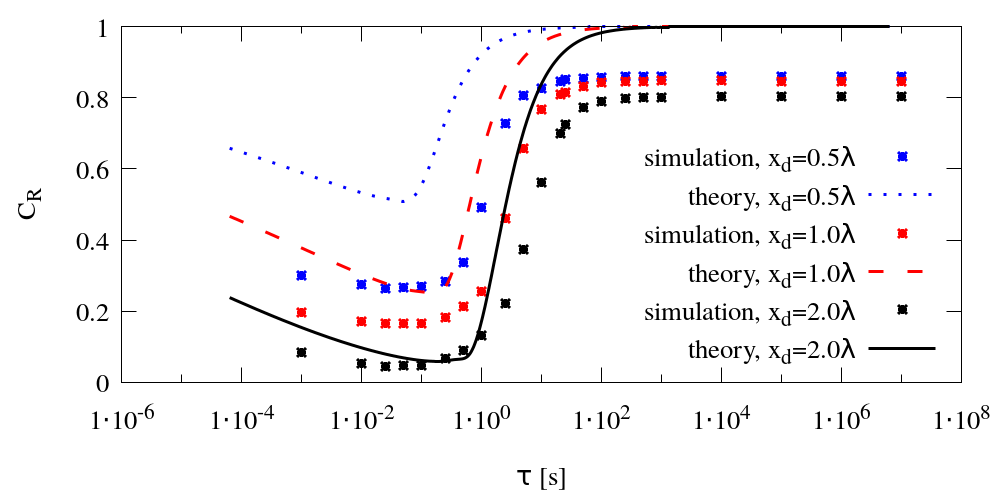} \\
$n = 2.8$ \\ 
 \includegraphics[width=\linewidth]{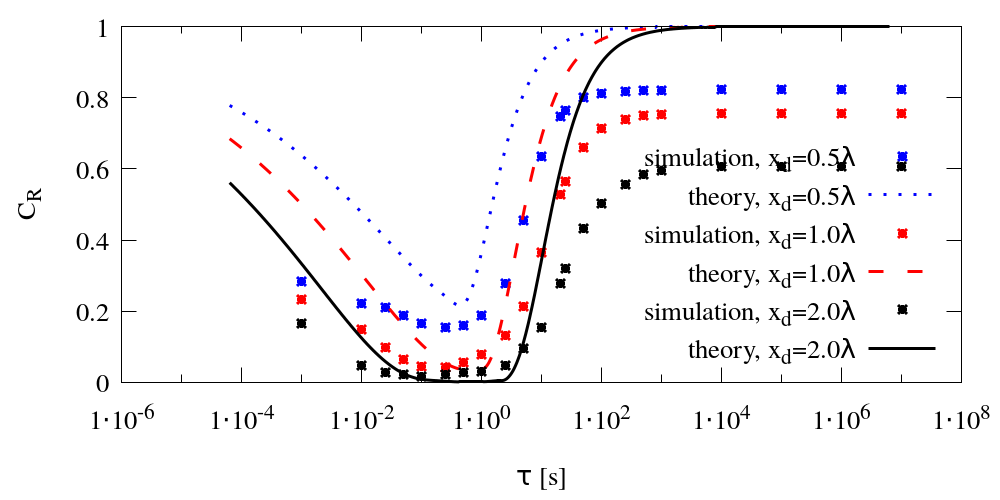} \\ 
$n = 0.46$ \\
\includegraphics[width=\linewidth]{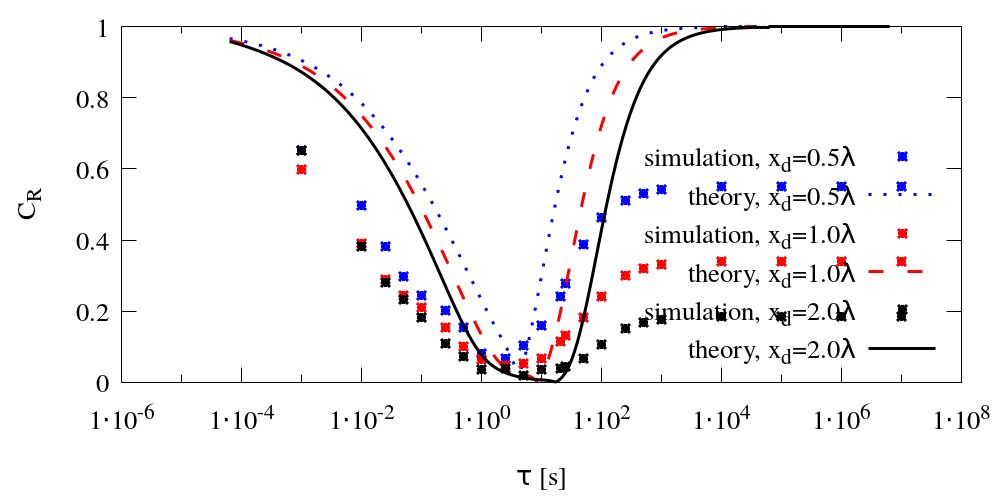} \\
$n = 0.1$ \\ 
 \includegraphics[width=\linewidth]{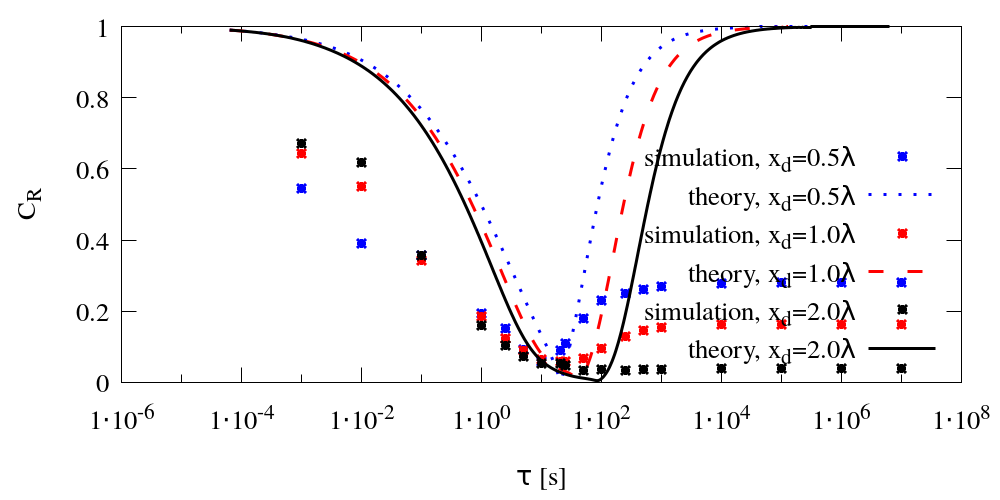} 
\end{center}
\end{small}
\caption{As Fig. \ref{FIG2Dexpn3p5coarseDeep}, except for power blending via Eq. (\ref{EQblendpow}) with different values for exponent $n$; the analytical approach (`theory') predicts the shift in optimum value for $\tau$ when changing exponent $n$; for all simulation results $C_{\mathrm{R,sim}}$ and corresponding analytical predictions $C_{\mathrm{R,theory}}$ holds $C_{\mathrm{R,sim}} - C_{\mathrm{R,theory}} < 8.4\%$ ($n=0.1$), $<4.7\%$ ($n=0.46$), $<4.2\%$ ($n=2.8$), and $<2.3\%$ ($n=10$); for the forcing strength $\tau \leq \tau_{\mathrm{opt,theory}}$ closest to the theoretical optimum value $\tau_{\mathrm{opt,theory}}$ holds $C_{\mathrm{R,sim}} - C_{\mathrm{R,theory}} < 2.7\%$ ($n=0.1$), $<3\%$ ($n=0.46$), $<2.0\%$ ($n=2.8$), and $<-1.5\%$ ($n=10$)} \label{FIG2DpowerNcoarseDeep}
\end{figure}

Figure \ref{FIG2DpowerNcoarseDeep} shows results for power blending according to Eq. (\ref{EQblendpow}) with different values of relaxation parameter $\tau$,  zone thickness $x_{\mathrm{d}}$, and  coefficient $n$. 
The results demonstrate that, depending on the choice of these parameters, the optimum value for relaxation parameter $\tau$ can vary by three orders of magnitude, which underlines the importance of optimizing the relaxation zone's parameters.
As before, the optimum value for $\tau$ is well predicted by the analytical approach. 

For $\tau\rightarrow \infty$, the relaxation source terms vanish to zero, so one would  expect that the solution behaves as if there were no relaxation zone; this would result in a standing wave (i.e. $C_{\mathrm{R}} \approx 1$), since the outlet boundary is nearly perfectly reflecting. Instead, for large $\tau$-values the reflection coefficients $C_{\mathrm{R}}$ were significantly lower than $1$, with lower values for smaller values of $n$. It is possible that this is due to the term $(1-b(\mathbf{x}))$ on the left-hand side of the governing equations: If there is no  reference solution to blend over to, then the blending out of the flow solution may behave like a damping. Note though that such large $\tau$-values are not of practical interest, because they cannot be used for combined generation and damping of waves as is illustrated in Sect. \ref{SECfsrelaxres3d} in Figs. \ref{FIG3Dmed} and \ref{FIG3Dfsfromvids}.

\begin{figure}[h!]
\begin{small}
\begin{center}
$x_{\mathrm{d}} = 0.5\lambda$ \\
\includegraphics[width=\linewidth]{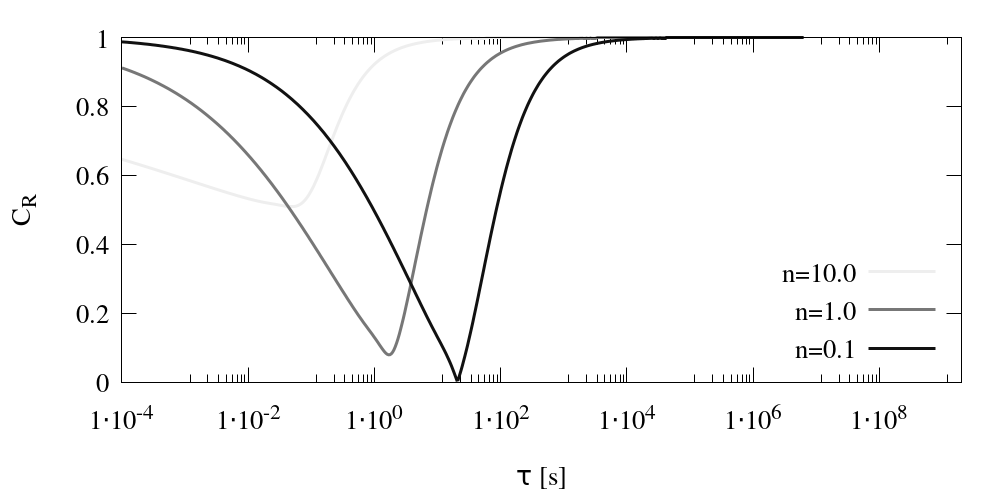} \\
$x_{\mathrm{d}} = 1\lambda$ \\ 
 \includegraphics[width=\linewidth]{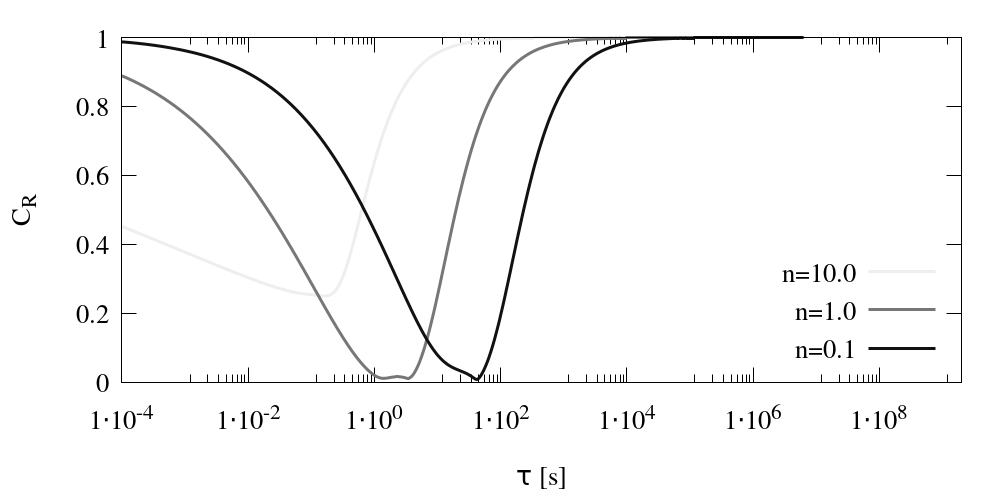} \\
$x_{\mathrm{d}} = 2\lambda$  \\ 
\includegraphics[width=\linewidth]{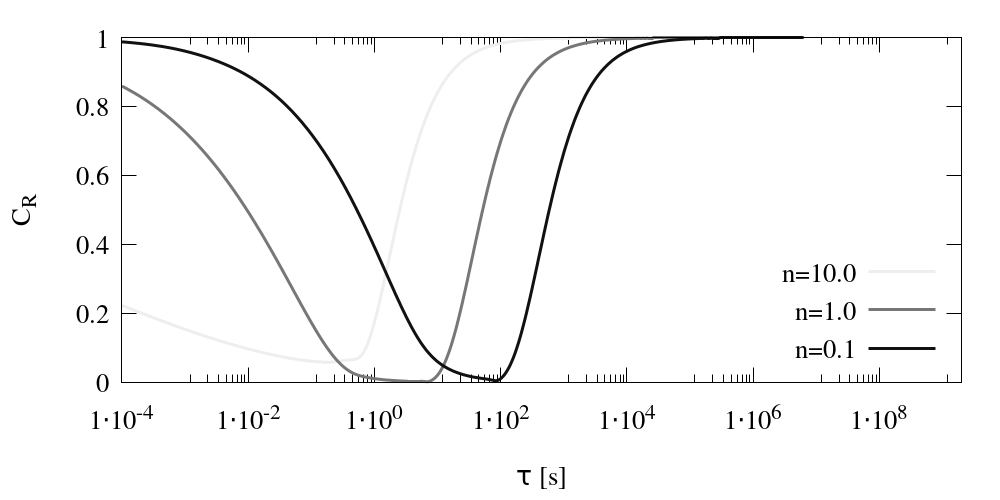} 
\end{center}
\end{small}
\caption{Analytical predictions (`theory') for reflection coefficient $C_{\mathrm{R}}$ as a function of relaxation parameter $\tau$ for deep-water waves with period $T=1.6\, \mathrm{s}$; for power blending according to Eq. (\ref{EQblendpow}) with different values for exponent $n$} \label{FIG2DpowerNcoarseDeeptheory}
\end{figure}

\begin{figure}[h!]
\begin{small}
\begin{center}
$x_{\mathrm{d}} = 0.5\lambda$\\
\includegraphics[width=\linewidth]{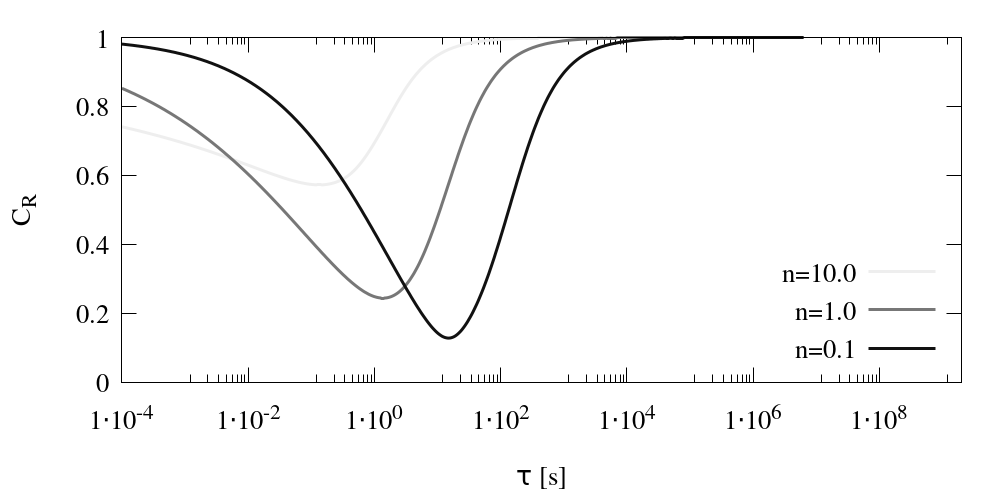} \\
$x_{\mathrm{d}} = 1\lambda$ \\ 
 \includegraphics[width=\linewidth]{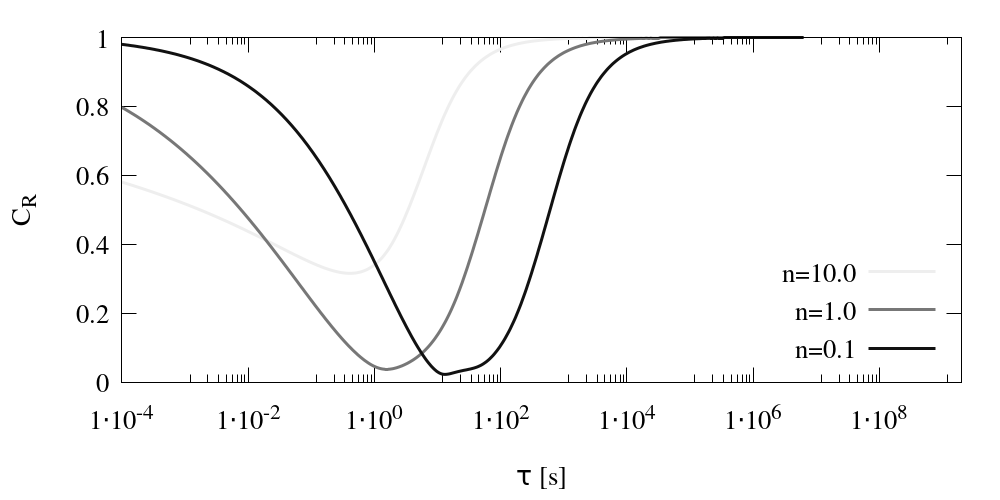} \\ 
$x_{\mathrm{d}} = 2\lambda$   \\ 
\includegraphics[width=\linewidth]{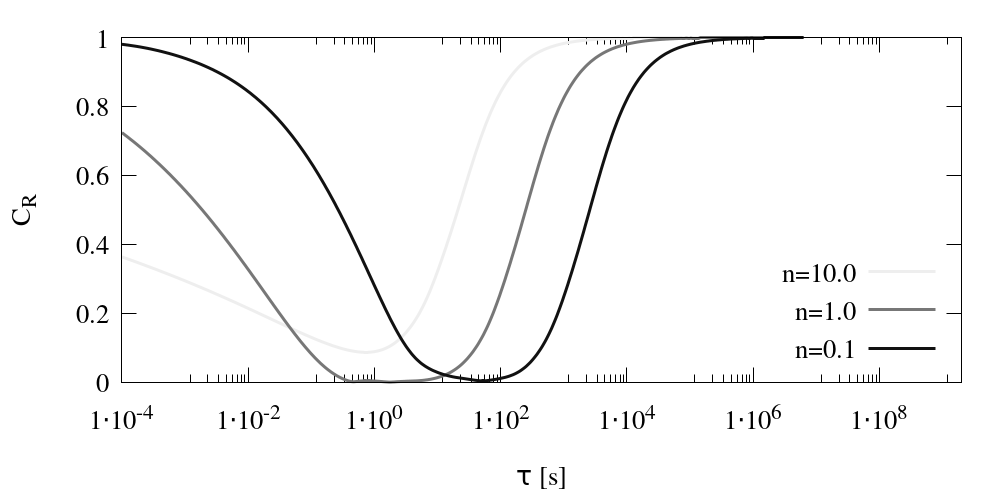} 
\end{center}
\end{small}
\caption{As Fig. \ref{FIG2DpowerNcoarseDeeptheory}, except for $\cos^{2n}$-blending according to Eq. (\ref{EQblendcos2}) with different values for exponent $n$; the optimum exponent $n$ increases with increasing zone thickness $x_{\mathrm{d}}$
} \label{FIG2DcosNcoarseDeeptheory}
\end{figure}

Not only do the optimum values for $\tau$ and  the curves for reflection coefficient $C_{\mathrm{R}}(\tau)$ change as a function of the blending function $b(\mathbf{x})$ (cf. Fig. \ref{FIG2DpowerNcoarseDeep}), but also the optimum choice of blending function (or here: its coefficient $n$) depends on the zone thickness $x_{\mathrm{d}}$ as Figs. \ref{FIG2DpowerNcoarseDeeptheory} and \ref{FIG2DcosNcoarseDeeptheory} demonstrate. The optimum choice of $n$ would correspond to the setting that  provides both the lowest reflection coefficient $C_{\mathrm{R}}$ for optimized $\tau$ and  the broadest range of adjacent $\tau$-values, for which the reflection coefficient $C_{\mathrm{R}}$ will be below a given threshold; the broader this range, the less sensitive will the reflection behavior of the relaxation zone be to changes of the wave period. Thus, irregular waves with a broad-banded wave energy spectrum can require a different (possibly larger) $n$-value than monochromatic waves.

For the investigated blending functions in Figs. \ref{FIG2DpowerNcoarseDeeptheory} and \ref{FIG2DcosNcoarseDeeptheory}, the larger the relaxation zone thickness $x_{\mathrm{d}}$ becomes, the larger becomes the optimum value for $n$. For practical choices of $x_{\mathrm{d}}$, the tendency appears to be that  $n$ should be $<1$ for $x_{\mathrm{d}}\lesssim 1.0\lambda$ and that $n$ should be $>1$ for $x_{\mathrm{d}}\gtrsim 1.5\lambda$. 

\section{Results from 3D-flow simulations}
\label{SECfsrelaxres3d}
To investigate the validity of the present findings for practical 3D-flow simulations, the flow around a strongly reflecting semi-submerged pontoon subjected to steep deep-water waves is simulated with the setup from Sect. \ref{SECfsrelaxsetup}. The solution domain was selected intentionally small, with implicit relaxation zones attached to all vertical domain boundaries with a zone thickness of only $x_{\mathrm{d}}\approx 0.7\lambda$. With respect to the tuning for the optimum blending function from  Sect. \ref{SECfsrelaxchoice0blending}, a power blending according to Eq. (\ref{EQblendpow}) with coefficient $n=0.46$ was used. According to the analytical prediction, a relaxation parameter of $\tau=2.5\, \mathrm{s}$ should provide a satisfactory reduction of undesired wave reflections ($C_{\mathrm{R}}\lesssim 5\%$), whereas $\tau$-values larger or smaller by a factor of $10^{\pm 1}$ should produce significant reflections. 

\begin{figure}[h!]
\begin{center}
\includegraphics[width=\linewidth]{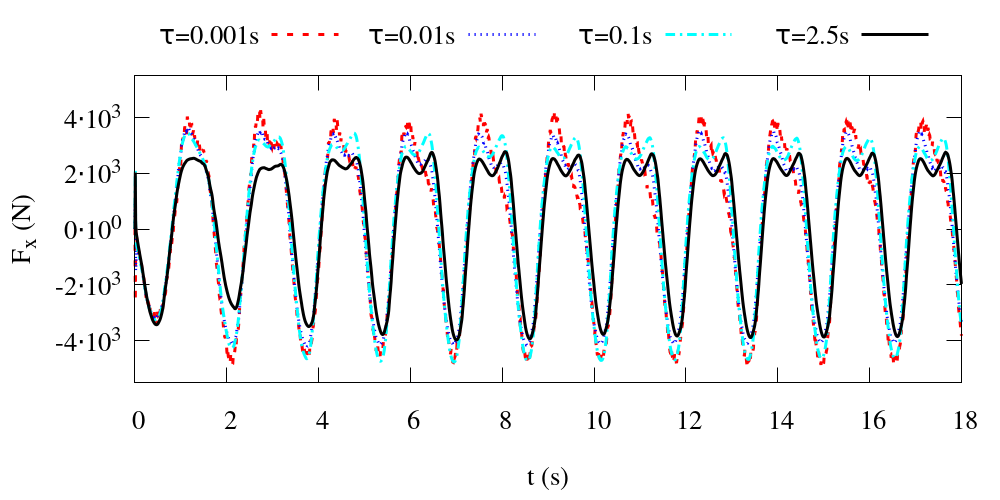}
\includegraphics[width=\linewidth]{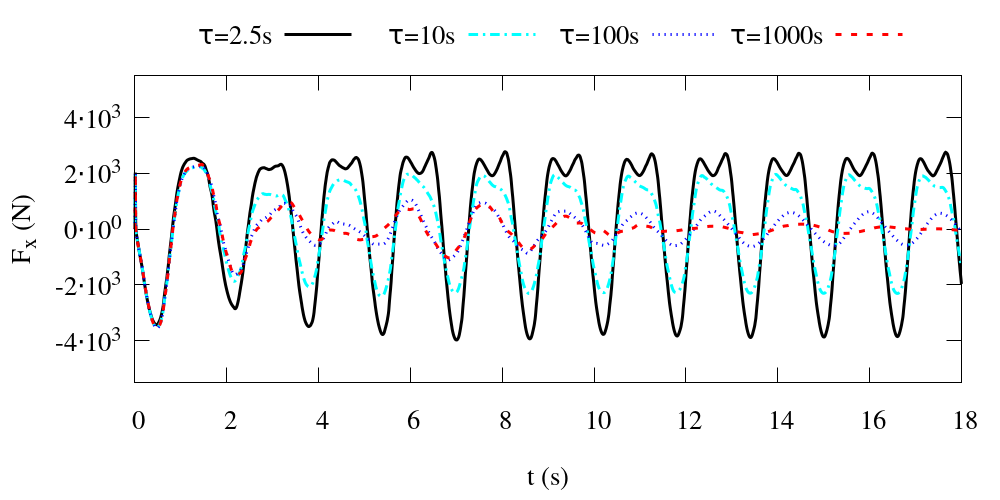}\\
\end{center}
\caption{Force component in $x$-direction integrated over the pontoon surface as a function of time $t$  for the \textit{medium grid}; with relaxation zone thickness $x_{\mathrm{d}}\approx 0.7\lambda$, power blending with exponent $n=0.46$ and different values of relaxation parameter $\tau$; the analytical approach from Sect. \ref{SECtheory} predicts an optimum of $\tau=2.5\, \mathrm{s}$, for which a periodic solution is obtained; the further $\tau$ deviates from the theoretical optimum, the stronger are the visible influences of undesired wave reflections in the results} \label{FIG3Dmed}
\end{figure}

If reflections are satisfactorily reduced, then a periodic solution is expected to occur after several wave periods, and long-time simulations should be possible without the accumulation of errors due to undesired wave reflections. Figure \ref{FIG3Dmed} shows that indeed such periodic results are obtained for the optimum setting. 

\begin{figure}[h!]
\begin{center}
\includegraphics[width=\linewidth]{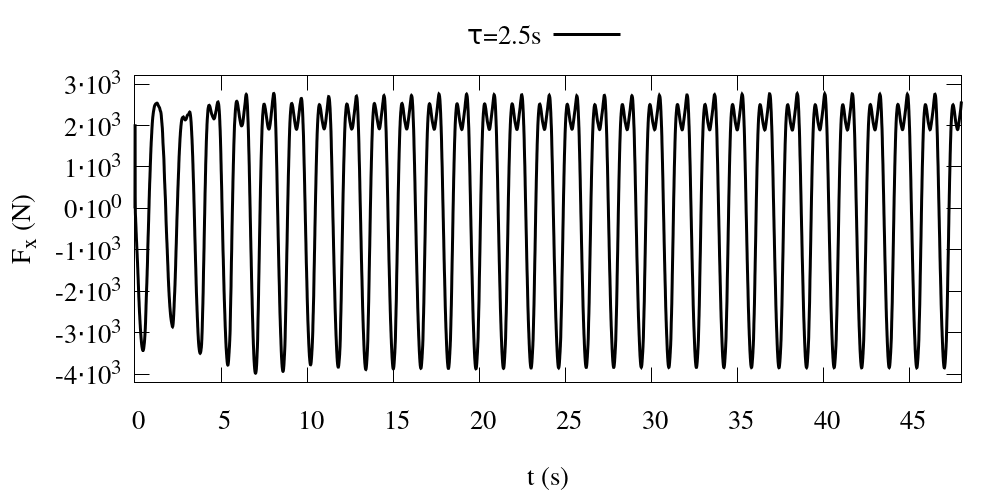}\\
\includegraphics[width=\linewidth]{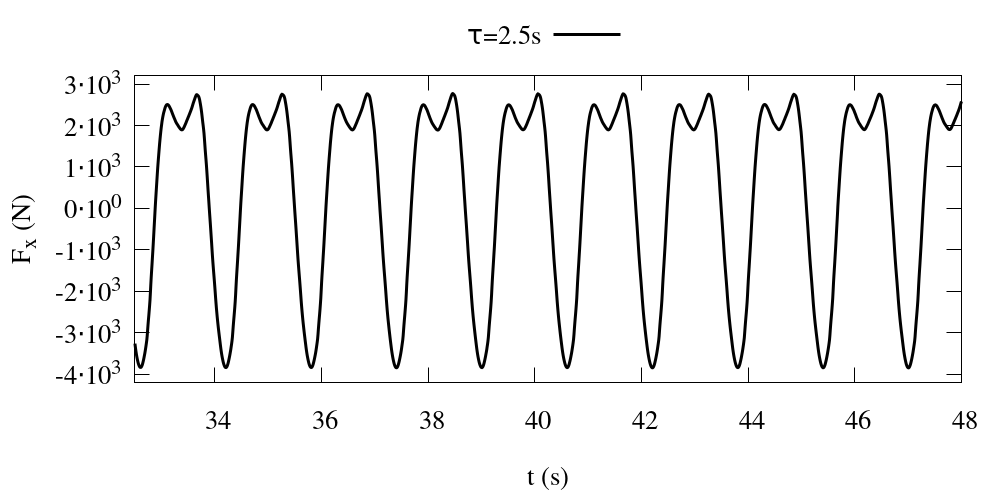}
\end{center}
\caption{As Fig. \ref{FIG3Dmed}, except for longer simulation duration; with  close-to-optimum relaxation  ($\tau=2.5\, \mathrm{s}$), simulations were performed for $30T$ without noticeable accumulation of reflections} \label{FIG3Dcoarselong}
\end{figure}

Figure \ref{FIG3Dcoarselong} shows that the correct tuning of the  relaxation zone enables a periodic solution for simulations over arbitrarily long simulation times. That the analytical solution from Eqs. (\ref{EQnavier_stokes})-(\ref{EQrelaxGamma}) applies with good approximation to the 3D-case as well has been shown both analytically and via 3D-flow simulations for a comparable case in Perić and Abdel-Maksoud (2020,2021) and is not repeated here.

\begin{figure}[h!]
\begin{small}
\begin{center}
too-strong relaxation ($\tau = 0.01\, \mathrm{s}$)  \\ 
\includegraphics[width=\linewidth]{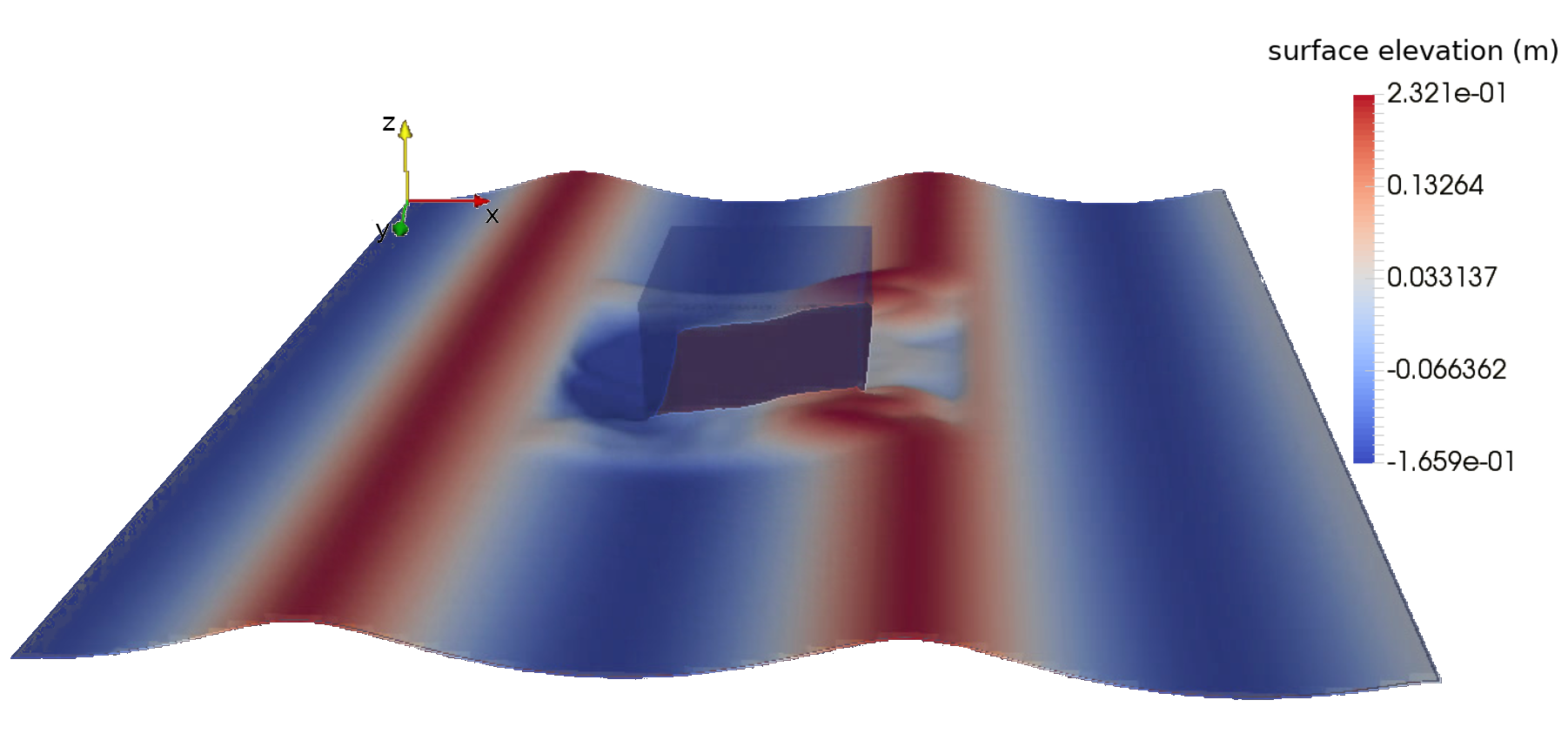} 
close-to-optimum relaxation ($\tau = 2.5\, \mathrm{s}$) \\
 \includegraphics[width=\linewidth]{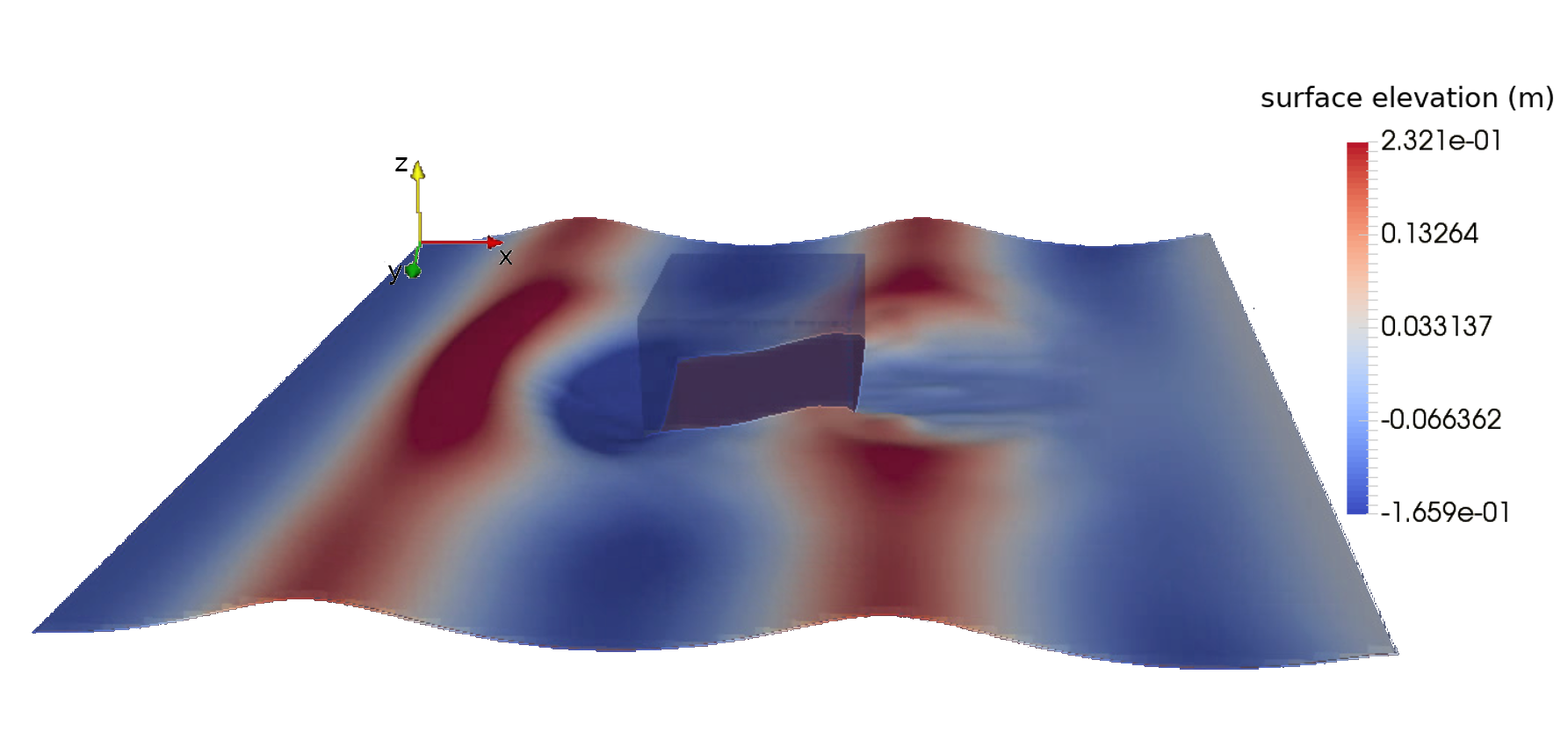} \\
too-weak relaxation ($\tau = 100\, \mathrm{s}$)   \\ 
\includegraphics[width=\linewidth]{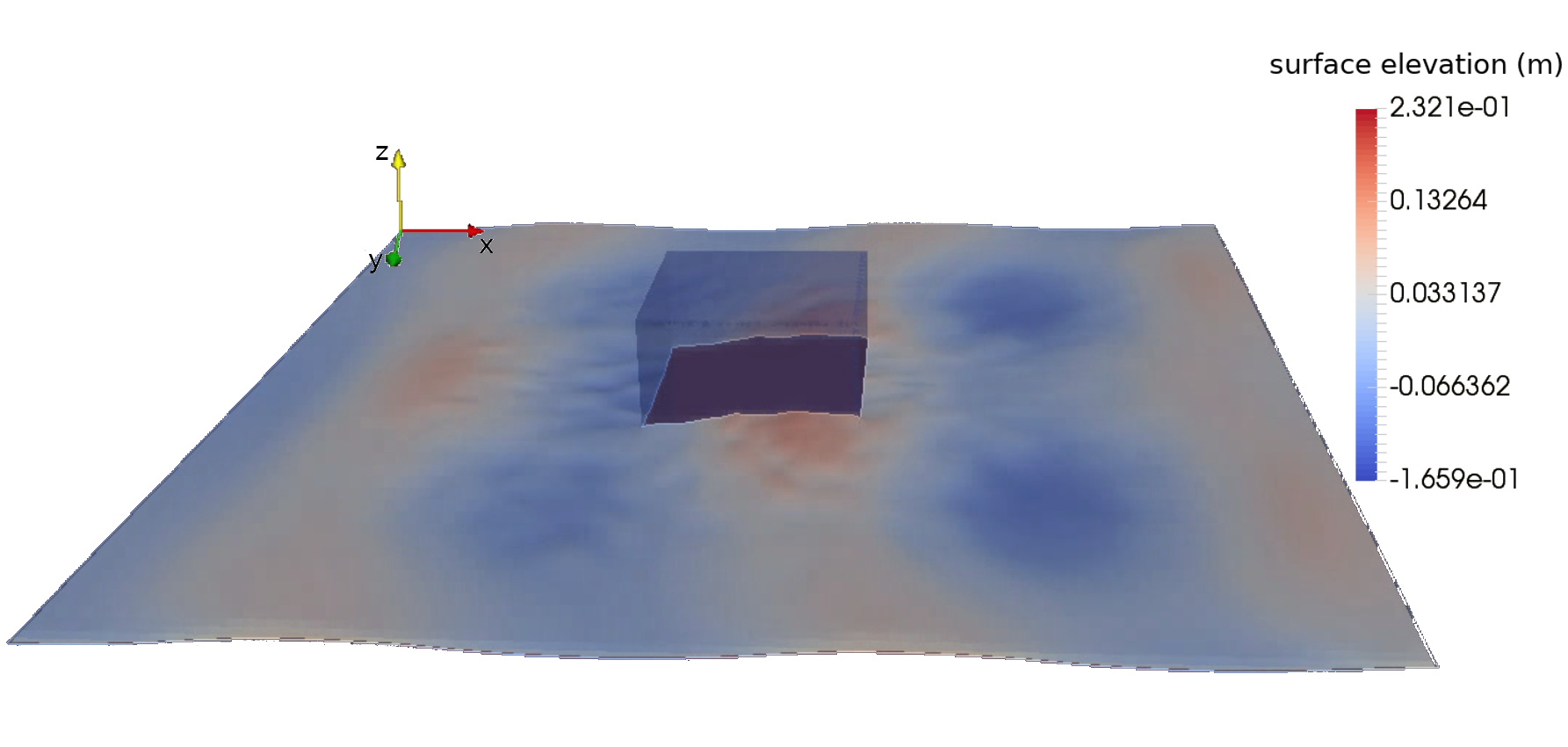}  
\end{center}
\end{small}
\caption{Simulation results for free-surface elevation at $t \approx 15\, \mathrm{s}$ for different values of relaxation parameter $\tau$; for relaxation-zone thickness $x_{\mathrm{d}}\approx 0.7\lambda$ and blending via Eq. (\ref{EQblendpow}) with exponent $n=0.46$; if the relaxation is too strong (top), wave-reflection occurs near the entrance to the relaxation zone; if relaxation is too weak (bottom), the far-field wave is not sustained; for optimized relaxation-setup (middle), the waves reflected at the pontoon decay smoothly over the whole relaxation zone as intended
} \label{FIG3Dfsfromvids}
\end{figure}

As shown in Fig. \ref{FIG3Dfsfromvids}, too-strong relaxation (corresponding to smaller-than-optimum $\tau$-values) produces wave reflections mainly at the entrance to the relaxation zone, resulting in  a change of amplitude as well as aperiodicity of the forces on the pontoon  (cf. Fig. \ref{FIG3Dmed}, upper plot). Too-weak relaxation  (corresponding to larger-than-optimum $\tau$-values) damps not only the undesired wave reflections, but also the incident wave,  so that the far-field wave is not sustained anymore, resulting in too-low forces on the pontoon (cf. Fig. \ref{FIG3Dmed}, bottom plot). 

\begin{figure}[h!]
\begin{small}
\begin{center}
coarse discretization \\ 
\includegraphics[width=\linewidth]{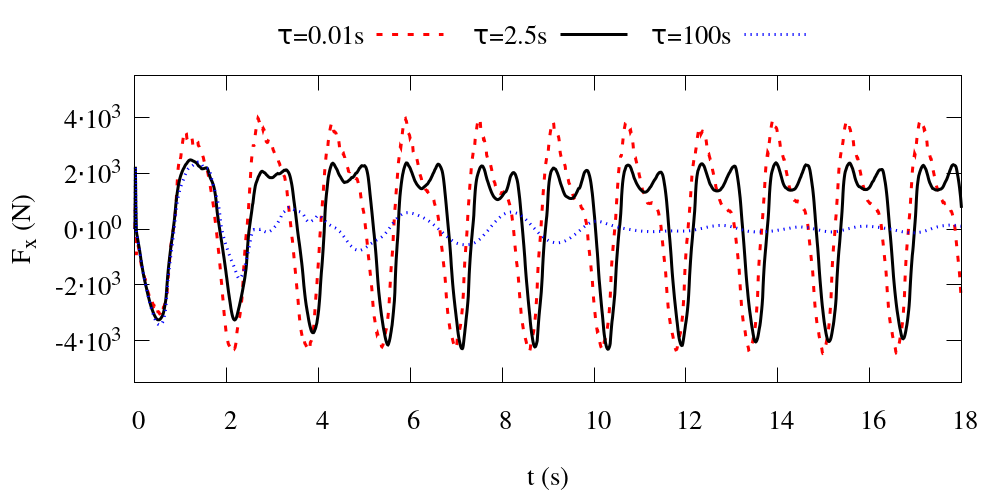} 
medium discretization \\
 \includegraphics[width=\linewidth]{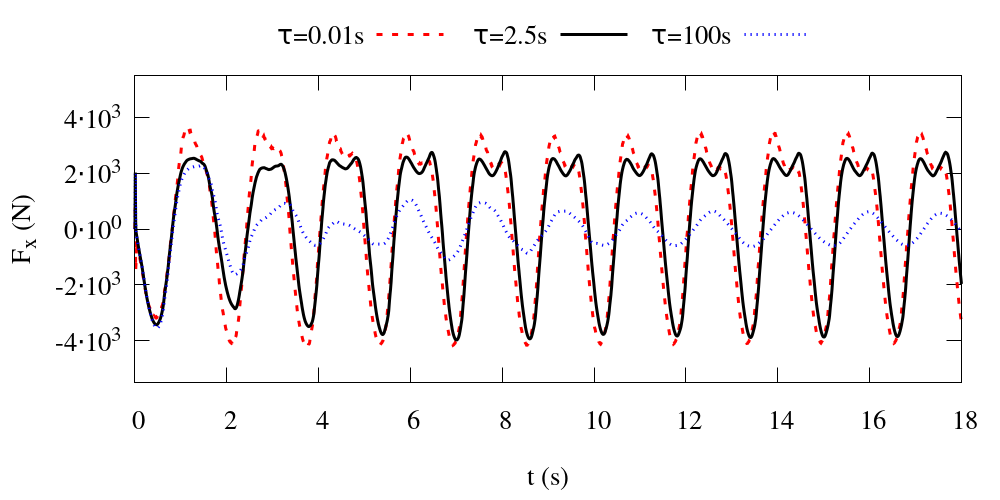} \\ 
fine discretization   \\ 
\includegraphics[width=\linewidth]{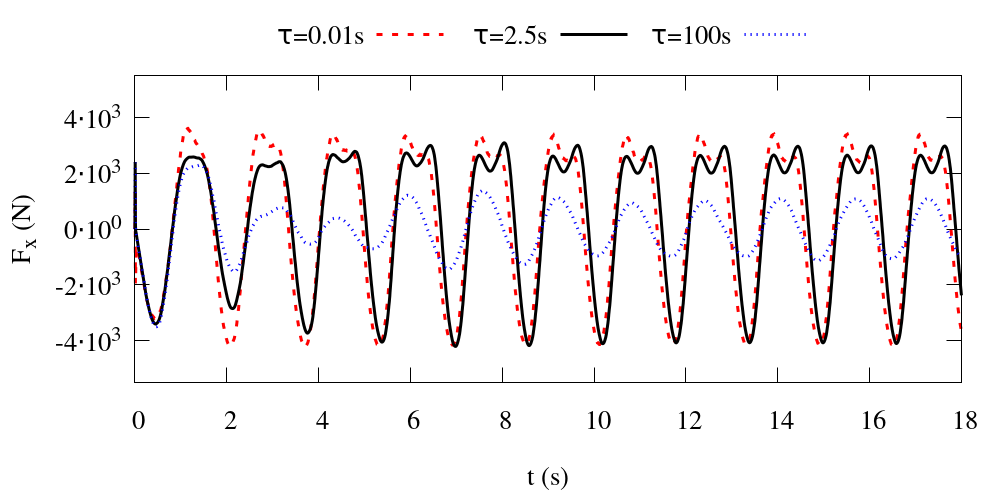}  
\end{center}
\end{small}
\caption{As Fig. \ref{FIG3Dmed}, except for coarse, medium and fine discretization} \label{FIG3Dgridstudy}
\end{figure}

\begin{figure}[h!]
\begin{center}
\includegraphics[width=\linewidth]{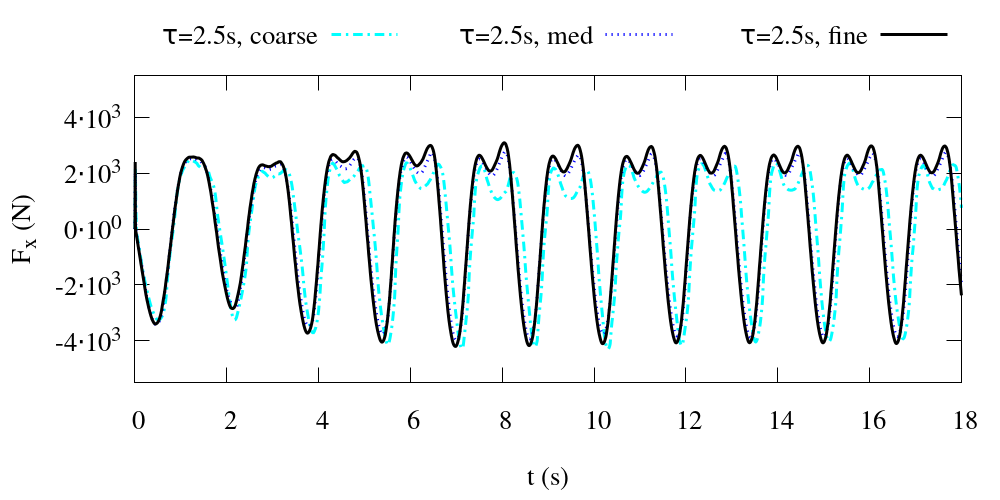}\\
\includegraphics[width=\linewidth]{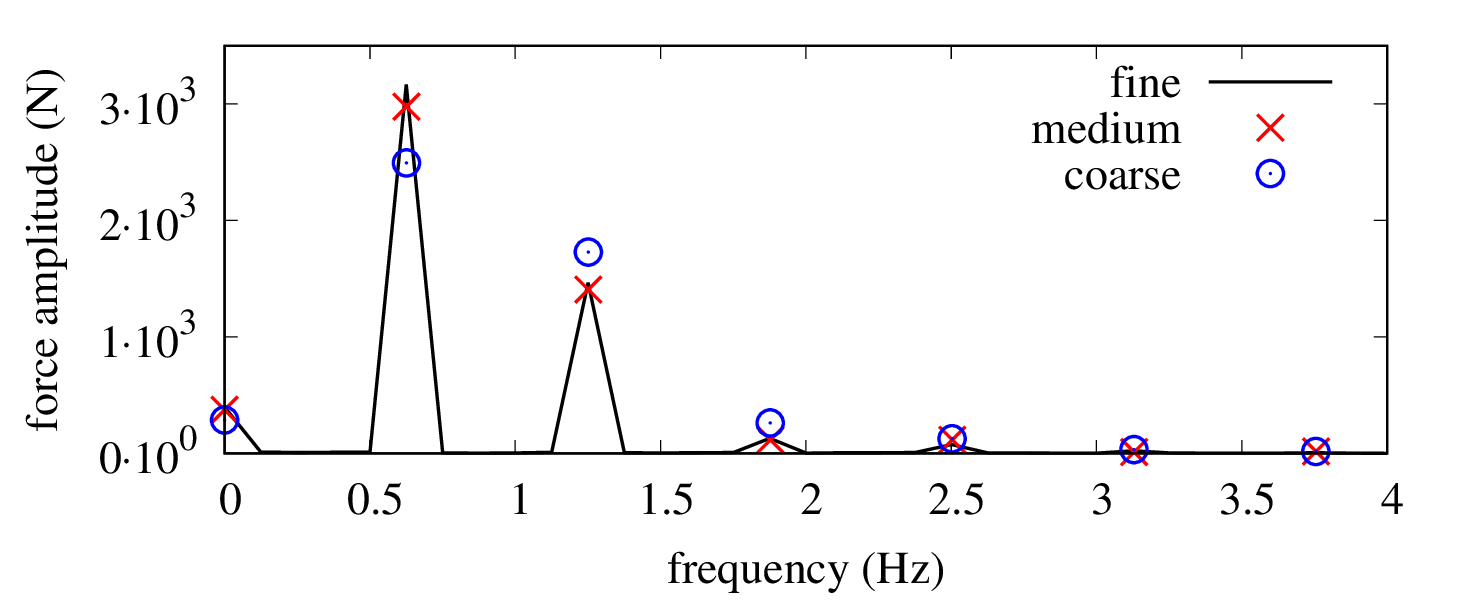}
\end{center}
\caption{Top: As Fig. \ref{FIG3Dmed}, except for close-to-optimum relaxation parameter $\tau$ on coarse, medium and fine discretization; bottom: force amplitude as a function of frequency for FFT-analysis of the curves from the upper plot during  time interval $10\, \mathrm{s}\leq t \leq 18\, \mathrm{s}$; between the coarse and fine discretization, all differences are $<660\, \mathrm{N}$, which corresponds to over $ 20\%$ of the first harmonic's amplitude; between the medium and fine discretization,all differences are $<190\, \mathrm{N}$, which corresponds to $\approx 6\%$ of the first harmonic's amplitude, so already the medium discretization is considered acceptable for the present purposes} \label{FIG3Dgridstudy2}
\end{figure}

Figures \ref{FIG3Dgridstudy} and \ref{FIG3Dgridstudy2} show that the difference between medium and fine grid is comparatively small, but for the coarse grid the force amplitudes are ca. $ 10\%$ lower. For the present purposes, all grids were considered suitable to demonstrate the benefits of tuning relaxation zones to the wave parameters.

\section{Discussion}
\label{SECdiscuss}
The results from Sects. \ref{SEC2dflow} and \ref{SECfsrelaxres3d} show that the analytical approach presented in Sect. \ref{SECtheory} is suitable for optimizing the case-dependent parameters of implicit relaxation zones, both for 2D- and complex 3D-flow simulations with nonlinear free-surface waves. When the implicit relaxation zones were optimized according to the analytical approach, the simulation results for reflection coefficient $C_{\mathrm{R}}$ were mostly lower or equal their analytical predictions, but never more than $3.4\%$ larger.

The analytical approach from Sect. \ref{SECtheory} closely predicted the optimum relaxation-zone parameters. For all simulation results, the optimum value for relaxation parameter $\tau$  was within $[\frac{1}{12}\tau_{\mathrm{opt,theory}},2\tau_{\mathrm{opt,theory}}]$ in terms of the analytically predicted optimum $\tau_{\mathrm{opt,theory}}$. 

It was found that simulation results for reflection coefficient $C_{\mathrm{R}}$ can be lower than predicted analytically, which occurred especially for smaller-than-optimum values of relaxation parameter $\tau$. The reason for this is that the analytical approach neglects that wave reflections due to source terms in different governing equations can have different phases and thus can partially cancel due to destructive interference. Future work will focus on considering this effect in the analytical approach, to obtain more accurate predictions of $C_{\mathrm{R}}$. 

However, already in its present form, the analytical approach predicts the relevant flow features and is sufficiently accurate to optimize the implicit relaxation zone's parameters. The analytical prediction for $C_{\mathrm{R}}$ can be considered as a close estimate of the upper-limit of the reflection coefficient $C_{\mathrm{R}}$ that will occur in the flow simulations.

How does optimizing the relaxation zone's parameters compare to using the default settings? The default value for $\tau$ in the Naval Hydro Pack is the time-step, i.e. $\tau = \Delta t$. This is a better choice than setting $\tau$ to a constant value, because $\tau = \Delta t$ scales correctly from model to full scale and produces a favorable matrix conditioning. However, it does not coincide with the optimum $\tau$-value: In Fig. \ref{FIG2Dexpn3p5coarseDeep}   $\tau=\Delta t$ is  up to two and in Fig. \ref{FIG2DpowerNcoarseDeeptheory} even up to four orders of magnitude smaller than optimal.
Moreover, reflection  increases if the time-step is refined, thus in a discretization-dependence study the results may not converge. It is therefore both more effective and more reliable to optimize the relaxation zone's parameters.

Optimizing the relaxation zone's parameters $\tau$, $x_{\mathrm{d}}$ and $b(\mathbf{x})$ also enables the use of thinner relaxation zones.  
With correct optimization, already  a zone thickness  of $0.5\lambda \leq x_{\mathrm{d}} \leq 1.0\lambda$ (depending on the intended reflection coefficient $C_{\mathrm{R}}$) minimizes undesired wave reflections satisfactorily. With default settings, an at least two to three times larger zone thickness  would be required to obtain the same reduction of undesired reflections (cf. Sect. \ref{SECfsrelaxchoice0blending}). Considering that typical values for zone thickness $x_{\mathrm{d}}$ in literature are $1\lambda \leq x_{\mathrm{d}} \leq 4\lambda$ (cf. Chen et al., 2019),   the computational effort can typically be reduced significantly when the relaxation zone's parameters are optimized.

The necessity of optimizing the case-dependent parameters of implicit relaxation zones becomes apparent when considering that, from short ocean waves to tidal waves, the wave period $T$ and correspondingly the optimum value for relaxation parameter $\tau$ can vary by factor $10^{4}$ or more, while variation of zone thickness $x_{\mathrm{d}}$ and blending function $b(\mathbf{x})$ can introduce a further variation by factor of $10^{3}$ or more. Consequently, the optimum $\tau$-value can vary by $7$ orders of magnitude for various marine applications.


It was shown that implicit relaxation zones can  be considered as a special-case of forcing zones. Therefore, findings obtained for forcing zones can be applied to implicit relaxation zones and vice versa, following the procedure outlined in this work. This is supported by the findings from Sect. \ref{SECfsrelax2Dstarfoamcomp}, which indicate that, when correctly optimized, forcing zones and implicit relaxation zones produce similar reduction of undesired wave reflections. Thus, the present results do not point in favor of one method over the other, rather the method available in one's flow solver should be used. 

Future research will focus on extending the analytical approach from Perić and Abdel-Maksoud (2018) to explicit relaxation zones. If this is achieved,  a unified formulation for the analytical description of all approaches for wave-generation and wave-damping based on domain-internal source terms has been obtained.

\section{Conclusion}
\label{SECconclusion}
An analytical approach was proposed for optimizing the case-dependent parameters of  implicit relaxation zones before performing the flow simulations. A computer program that evaluates the analytical approach has been published as free software, to facilitate use of the approach in engineering practice.

The analytical predictions were validated against flow simulation results using two different codes, the \texttt{foam-extend} Naval Hydro Pack and Siemens STAR-CCM+. Flow simulations of free-surface wave propagation with implicit relaxation zones were performed or a wide range of settings for the relaxation zone's parameters, for relaxation towards different reference solutions, under shallow-water and deep-water conditions, for different wave periods,  for both linear and nonlinear waves with up to $66\%$ of breaking steepness, and for both 2D- and complex 3D-flow problems.

The analytical predictions for the optimum values of the case-dependent parameters matched closely with the corresponding simulation results. When the implicit relaxation zones were optimized as proposed, the simulation results for reflection coefficient $C_{\mathrm{R}}$ were mostly lower or equal their theoretical predictions, but never more than $3.4\%$ larger.

Furthermore, it was demonstrated that optimizing the relaxation zone's parameters  enables the use of significantly thinner zones and thus reduces the computational effort.  Therefore, the proposed analytical approach can be recommended for optimizing implicit relaxation zones in engineering practice.

\section*{Acknowledgements}
The study was supported by the Deutsche Forschungsgemeinschaft (DFG) with grants AB 112/11-1 and AB 112/11-2.


\end{document}